\documentclass[prl,aps,twocolumn,superscriptaddress]{revtex4-1}
\usepackage[utf8]{inputenc}
\usepackage{amsmath, amssymb, amsbsy}
\usepackage{color}
\usepackage{array}
\usepackage{bm, bbm}
\usepackage{graphicx}
\usepackage{diagbox, multirow}
\usepackage[percent]{overpic}
\usepackage{epsfig}
\usepackage{extarrows}
\usepackage{appendix}
\usepackage[normalem]{ulem}
\usepackage{xcolor}
\usepackage{braket}
\usepackage{tikz}
\usetikzlibrary{tikzmark}
\usepackage{xcolor}

\usepackage[colorlinks=true,linkcolor=blue,citecolor=blue,urlcolor=blue,bookmarks=false]{hyperref}

\pagecolor{white}

\newcommand{\1}{\hspace*{-1pt}}
\newcommand{\2}{\hspace*{-2pt}}
\newcommand{\3}{\hspace*{-3pt}}

\begin{document}

\title{Bilinear Majorana representations for spin operators with spin magnitudes $S>1/2$}

\date{\today}

\author{Yannik Schaden}
\affiliation{Dahlem Center for Complex Quantum Systems and Institut f\"ur Theoretische Physik, Freie Universit\"at Berlin, Arnimallee 14, 14195 Berlin, Germany}
\affiliation{Helmholtz-Zentrum Berlin f\"ur Materialien und Energie, Hahn-Meitner-Platz 1, 14109 Berlin, Germany}

\author{Johannes Reuther}
\affiliation{Dahlem Center for Complex Quantum Systems and Institut f\"ur Theoretische Physik, Freie Universit\"at Berlin, Arnimallee 14, 14195 Berlin, Germany}
\affiliation{Helmholtz-Zentrum Berlin f\"ur Materialien und Energie, Hahn-Meitner-Platz 1, 14109 Berlin, Germany}
\affiliation{Department of Physics and Quantum Centers in Diamond and Emerging Materials (QuCenDiEM) group, Indian Institute of Technology Madras, Chennai 600036, India}

\begin{abstract}
We present a classification of bilinear Majorana representations for spin-$S$ operators, based on the real irreducible matrix representations of SU(2). We identify two types of such representations: While the first type can be straightforwardly mapped onto standard complex fermionic representations of spin-$S$ operators, the second type realizes spin amplitudes $S=s(s+1)/4$ with $s\in\mathbb{N}$ and can be considered particularly efficient in representing spins via fermions. We show that for $s=1$ and $s=2$ this second type reproduces known spin-$1/2$ and spin-$3/2$ Majorana representations and we prove that these are the only bilinear Majorana representations that do not introduce any unphysical spin sectors. While for $s>2$, additional unphysical spin spaces are unavoidable they are less numerous than for more standard complex fermionic representations and carry comparatively small spin amplitudes. We apply our Majorana representations to exactly solvable small spin clusters and confirm that their low energy properties remain unaffected by unphysical spin sectors, making our representations useful for auxiliary-particle based methods.
\end{abstract}
\maketitle

\section{Introduction}
The theoretical investigation of quantum spin models has always constituted a particularly subtle subject in condensed matter physics, because their purely interacting nature leads to a lack of small interaction limits. Furthermore, the spin commutator algebra is more complicated than the corresponding relations for bosonic or fermionic operators. Since the early days of quantum mechanics, important progress in (approximately) solving these systems has often been made with the help of spin representations, where the spin operators are expressed in terms of auxiliary particles, either of bosonic or of fermionic type. There are numerous well-established ways for such rewritings of spin operators, each of which has its own characteristic range of application where it allows to gain insights into the physical properties of quantum spin systems. 

In a pioneering work by Jordan and Wigner~\cite{jordan28}, spin $S=1/2$ operators are expressed via fermions which is very useful for 1D spin systems but becomes more complicated in higher dimensions. On the bosonic side, the Holstein-Primakoff representation~\cite{holstein40} holds for arbitrary spin magnitudes $S$ and is usually applied in 2D or higher dimensions where it works best in the quasiclassical $S\rightarrow\infty$ limit. Particularly, it allows to describe small deviations from a classical magnetically ordered state via a $1/S$ expansion giving rise to spin wave excitations~\cite{anderson52,kubo52}. The opposite scenario where a spin system is strongly affected by quantum fluctuations hindering the formation of long-range magnetic order is known to be amenable to Abrikosov fermions~\cite{abrikosov_electron_1965}. This bilinear, fermionic rewriting of $S=1/2$ spin operators (which can also be generalized to larger $S$~\cite{zhang06,hong09,liu10}) directly incorporates the type of fractionalization of spinful quasiparticles that is deemed to be characteristic for quantum spin liquids~\cite{savary_quantum_2017}, and, hence, constitutes the basis for ``parton mean-field theories''~\cite{affleck88,wen91,wen02}. While the Abrikosov spin representation suffers from the existence of unphysical subspaces, it is exactly this disadvantage that also enables a very insightful description of quantum spin liquids in terms of emergent gauge theories~\cite{wen_book}. Notably, the Abrikosov representation also provides the starting point for numerical methods in quantum magnetism such as the pseudo fermion functional renormalization group (PFFRG)~\cite{reuther_j_2010,iqbal16,baez_numerical_2017,rueck18,roscher18,buessen18,buessen19,buessen_code,kiese20,thoenniss20,ritter22,kiese22,schneider22}. The bosonic (and again bilinear) counterpart of the Abrikosov representation is the Schwinger boson transformation which may, likewise, be used to explore quantum spin liquids via an approach called ``Schwinger-boson mean-field theory''~\cite{arovas88,auerbach_book,zhang22}. Since bosons can undergo condensation~\cite{hirsch89,sarker89,chandra90}, this technique describes spin liquid behavior and long-range magnetic order within the same formalism by associating them with gapped and gapless bosonic excitations, respectively. This, however, comes with the disadvantage that only gapped quantum spin liquids can be accessed.

In recent years, Majorana representations for $S=1/2$ spin operators have gained increasing popularity. The strength of such Majorana rewritings becomes most apparent when applying them to the Kitaev honeycomb model where they enable an exact analytic solution of the system's spin liquid ground state~\cite{kitaev_anyons_2006}. In addition, a Majorana formulation of the Kitaev honeycomb model uncovers a very elegant gauge structure where free fermions couple to a static $\mathbb{Z}_2$ gauge field. Majorana representations have also been extensively used in generalizations of the honeycomb Kitaev model, particularly, for other types of three-coordinated lattices~\cite{yao07,mandal09,hermanns14,hermanns15,obrien16,rachel16,casella22}. Three variants of Majorana representations for $S=1/2$ spin operators have previously been applied, all being bilinear in the Majorana operators: Kitaev's representation, initially used for the solution of the Kitaev honeycomb model~\cite{kitaev_anyons_2006}, as well as the so-called SO(3)~\cite{martin59,tsvelik92} and SO(4)~\cite{chen12} representations, where the latter is equivalent to the aforementioned complex fermionic Abrikosov representation~\cite{abrikosov_electron_1965}. The SO(3) representation stands out in this list as it only requires three instead of four Majorana operators and, most remarkably, does not introduce any unphysical spin states, see Refs.~\cite{shastry97,mao03,shnirman03,biswas_su2-invariant_2011,schad14,schad15,schad16,niggemann_frustrated_2021,niggemann_quantitative_2022} for previous applications. A Majorana representation for $S=3/2$ spin operators has also been frequently used~\cite{wang_z_2009,yao09,yao11,chua11,natori_chiral_2016,natori_dynamics_2017,natori18,carvalho18,de_farias_quadrupolar_2020,natori20,jin_unveiling_2022,PhysRevB.83.060407,PhysRevX.12.041029}.

This raises the question whether Majorana representations can be constructed for arbitrarily large spin magnitudes $S$. The existence of such representations would indicate that Majorana fermions are not only useful for the description of quantum spin liquids at small $S$ but also in the opposite $S\rightarrow\infty$ limit where classical spin physics prevails. Our motivation for investigating this question is partially driven by our efforts in developing new numerical many-body techniques that are flexible with respect to the spin magnitude $S$. Additional insights into spin representations with $S>1/2$ are particularly beneficial in the context of the aforementioned PFFRG and variants thereof since they fundamentally rely on auxiliary particle representations for spin operators. It has recently been demonstrated that these methods allow one to implement spin-$1/2$ degrees of freedom via the SO(3) Majorana representation, yielding an approach called pseudo Majorana functional renormalization group (PMFRG)~\cite{niggemann_frustrated_2021,niggemann_quantitative_2022}. Due to the absence of unphysical states in this representation, the PMFRG has proven very powerful in describing magnetically ordered and disordered phases in quantum spin-$1/2$ systems~\cite{niggemann_frustrated_2021,niggemann_quantitative_2022}. Hence, generalizations of Majorana spin representations for larger $S$ will have a direct use in our ongoing efforts in developing these methods further. 

In this paper, we present a full classification of all possible bilinear representations of spin operators in terms of Majorana fermions. Our main result is that Majorana spin representations exist for arbitrary spin magnitudes $S$, however, a particularly appealing class of Majorana representations occurs for $S=s(s+1)/4$ (where $s$ is a positive integer), for which we provide explicit analytical expressions. Part of this class are the known Majorana representations for $S=1/2$ and $S=3/2$ which do not involve any unphysical states. For larger $S$, the Majorana fermions generate additional spin spaces, but far less than for more standard complex fermionic representations~\cite{zhang06,hong09,liu10} and with comparatively small spin amplitudes. For example, we identify a spin-$3$ representation which only comes along with a rather gentle addition of a single spin-$0$ subspace. We further demonstrate that this spin gap of $\Delta S=3$ between the largest realized spin amplitude $S=s(s+1)/4$ and the second largest spin amplitude is characteristic for our Majorana representations. In systems of interacting spins, the spin gap can be expected to translate into an energetic difference between the ground states in the physical and unphysical sectors. We confirm this expectation for clusters of a few interacting spins, where this energy difference is indeed found to be a robust property over various cluster sizes and spin magnitudes. We therefore consider these Majorana representations useful for future developments of analytical and numerical approaches for spin systems with $S>1/2$. Given the widespread use of auxiliary particle representations for spin operators our findings may also be helpful beyond the field of quantum magnetism.

The rest of the paper is structured as follows: In the preparatory Sec.~\ref{sec:preriquisites}, we briefly review various known properties of spin operators and introduce the standard complex fermionic representations. In Sec.~\ref{majoconstruction}, we systematically construct and characterize all bilinear Majorana representations and discuss their relations with complex fermionic representations. Section~\ref{sec:examples} focuses more specifically on a few Majorana representations with the smallest spin magnitudes and associates them with Majorana representations from previous literature. This section also briefly discusses possible approaches for projecting out unwanted spin spaces. Finally, Sec.~\ref{sec:states} investigates the impact of additional spin spaces on small interacting spin clusters if such a projection is not performed. The paper ends with a summary and conclusion in Sec.~\ref{sec:summary}.

\section{Prerequisites: Spin operators and complex fermionic spin representations} \label{sec:preriquisites}
As a preparation for the forthcoming sections we first state some general properties of spin operators and introduce the known {\it complex} fermionic spin representations for general spin magnitudes $S\geq 1/2$~\cite{zhang06,hong09,liu10}.

\subsection{Properties of spin operators}
The fundamental condition, defining a spin operator ${\bm S}=(S^x,S^y,S^z)$ is the spin algebra relation
\begin{align}
	[S^\mu, S^\nu] = \mathrm{i} \epsilon^{\mu \nu \rho} S^\rho, \label{spinalgebra1}
\end{align}
with $\mu,\nu,\sigma\in\{x,y,z\}$. Furthermore, the spin magnitude $s$ which can takes values $s\in\{1/2,1,3/2,\ldots\}$ follows from
\begin{align}
{\bm S}^2=s(s+1)\mathbb{I}_{2s+1},\label{spinmagnitude}
\end{align}
where $\mathbb{I}_{2s+1}$ is the identity operator in the $2s+1$ dimensional spin space. (We note that here, $s$ can also be half-integer while for our Majorana representations in Sec.~\ref{sec:odd} below $s$ will be restricted to be integer.) The well known matrix representations of these spin operators are, up to a factor 1/2, just the Pauli matrices and their spin $s$ generalizations. We denote them as $K^\mu$ in the following and they are defined as
\begin{equation}
K^\mu_{mn}=\langle s,m|S^\mu|s,n\rangle, \label{eq:matrixrepn}
\end{equation}
where $|s,m\rangle$ are the eigenvectors of $S^z$ with eigenvalues $m\in\{s,s-1,...,-s\}$. For completeness, we state the explicit matrix forms of $K^\mu$. By definition, $K^z$ is diagonal:
\begin{equation}\label{eq:iz}
K^z_{mn}=m\delta_{mn}.
\end{equation} 
Furthermore, $K^x$ and $K^y$ are matrices with finite elements on the first off-diagonal only, i.e.
\begin{align}\label{eq:ixiy}
K^x_{m,m+1}&=K^x_{m+1,m}=\frac{1}{2}\sqrt{s(s+1)-m(m+1)},\notag\\
K^y_{m,m+1}&=-K^y_{m+1,m}=\frac{\mathrm{i}}{2}\sqrt{s(s+1)-m(m+1)},
\end{align}
with $m\in\{-s,\-s+1\ldots,s-1\}$ while all other elements vanish.

By construction, the matrices $K^\mu$ again fulfill the angular momentum algebra upon matrix multiplication,
\begin{align}
	[K^\mu,K^\nu] = \mathrm{i} \epsilon^{\mu \nu \rho} K^\rho. \label{eq:Kalgebra}
\end{align}
With this property ${\bm K}=(K^x,K^y,K^z)$ are $2s+1$-dimensional generators of a matrix representation of SU(2), whose elements are given by $e^{-\mathrm{i}\varphi {\bm n}\cdot{\bm K}}$, where $\varphi\in[0,2\pi)$ is the rotation angle and ${\bm n}$ with $|{\bm n}|=1$ is the rotation axis. Additionally, the matrix representations of SU(2) following from the $K^\mu$ matrices in Eqs.~(\ref{eq:iz}) and (\ref{eq:ixiy}) are irreducible and (up to unitary transformations) they are the only irreducible representations of SU(2) which implies that for each integer dimension $d=2s+1$ there exists exactly one such representation~\cite{group_theory}.

\subsection{Construction of complex fermionic spin representations}\label{sec:complex_fermions}

The Majorana representations constructed in the next section will have a close relation to the complex fermionic representations of spin operators, therefore, it is useful to introduce the latter first. The complex fermionic representation of a spin operator $\bm{S}$ makes use of $2s+1$ Fermi operators $f_m$ with $m\in\{-s,\cdots,s\}$ satisfying the anticommutation relations
\begin{align}
	\{f_m,f_n^\dagger\}=\delta_{mn}, \hspace{15pt}
	\{f_m,f_n\}=\{f_m^\dagger,f_n^\dagger\}=0,
\end{align}
and is given by~\cite{zhang06,hong09,liu10}
\begin{equation}
S^\mu= f_m^\dagger K_{mn}^\mu f_n.\label{eq:complex_rep1}
\end{equation}
Here $K_{mn}^\mu$ is the matrix representation of a spin operator with magnitude $s$ as given in Eq.~(\ref{eq:matrixrepn}). The fermionic operators and, hence, $S^\mu$ in Eq.~(\ref{eq:complex_rep1}) act on a $2^{2s+1}$-dimensional Fock space, where basis states can be labelled by the occupation numbers of the fermionic modes $m$,
\begin{equation}\label{eq:occupation}
N_m=f_m^\dagger f_m\in\{0,1\}.
\end{equation}

While the spin representation in Eq.~(\ref{eq:complex_rep1}) fulfills Eq.~(\ref{spinalgebra1}), it relaxes the condition in Eq.~(\ref{spinmagnitude}) in the sense that it generates a direct sum of spin degrees of freedom with different magnitudes $S$ forming a set of numbers $\mathcal{M}(s)$ where $S\in\mathcal{M}(s)$. In other words, ${\bm S}^2$ calculated from Eq.~(\ref{eq:complex_rep1}) has a block diagonal form, where each block separately satisfies Eq.~(\ref{spinmagnitude}) with possibly different spin magnitudes $S$. Note that here and in the following one needs to strictly distinguish between $s$ which is a spin quantum number that one can assign to the matrices $K^\mu$ via Eq.~(\ref{eq:matrixrepn}) and $S$ which are the actual spin amplitudes generated by the auxiliary particle representation for spin operators, in the present case Eq.~(\ref{eq:complex_rep1}). The generation of multiple spin spaces is a common property of auxiliary particle representations of spin operators and, hence, we will allow for this type of relaxation of Eq.~(\ref{spinmagnitude}) throughout this work.

One way of determining the set of spin quantum numbers $\mathcal{M}(s)$ realized for a given $s$ is to consider the eigenvalues $M^z$ of $S^z$ from Eq.~(\ref{eq:complex_rep1}). Since $K^z$ is diagonal [see Eq.~(\ref{eq:iz})] they easily follow from the occupation numbers $N_m$,
\begin{equation}\label{eq:mz}
 M^z\in \Bigg\{\sum_{m=-s}^{s} m N_m \Bigg\}\Bigg|_{\{N_m=0,1\}}.
\end{equation}
Here, the set of quantum numbers $M^z$ is formed by all combinatorial possibilities of independently choosing the individual occupation numbers $N_m$ with $m\in\{-s,\ldots,s\}$ as $0$ or $1$. In each $2S+1$-dimensional subspace of the full $2^{2s+1}$-dimensional Hilbert space that corresponds to a specific spin quantum number $S$, the eigenvalues $M^z$ of $S^z$ are given by consecutive numbers $S,S-1,\ldots,-S$. Hence, to find all realized quantum numbers $S\in\mathcal{M}(s)$ one needs to identify these sequences in the set of Eq.~(\ref{eq:mz}) in such a way that all numbers in the set belong to a sequence (for which there is always a unique solution). For example, for $s=1/2$ and $s=1$ one finds $\mathcal{M}(1/2)=\{1/2,0,0\}$ and $\mathcal{M}(1)=\{1,1,0,0\}$, respectively, where the first is the standard Abrikosov representation for spin-$1/2$ operators~\cite{abrikosov_electron_1965} which comes along with two extra and trivial spin-$0$ subspaces. All these subspaces for spin amplitudes $S$ correspond to a total particle number
\begin{equation}
    N_f=\sum_{m=-s}^{s} N_m= \sum_{m=-s}^{s} f^\dagger_m f_m.
\end{equation}
For example, subspaces with $N_f=1$ and $N_f=2s$ realize spins with amplitude $S=s$~\cite{liu10} while subspaces with $N_f=0$ and $N_f=2s+1$ correspond to trivial spin singlets, $S=0$. Note that subspaces with other particle numbers may also realize multiple different spin spaces $S$ including spaces with $S>s$. There exists a variety of methods to single out the subspace with the desired spin magnitude $S$ by enforcing a particle number constraint, either approximately~\cite{reuther_j_2010,baez_numerical_2017}, or exactly~\cite{popov88,prokofiev11}. In the case of the spin-$1/2$ Abrikosov representation an elegant and {\it exact} approach is to project out the spin-$0$ subspaces with the Popov-Fedotov method~\cite{popov88,prokofiev11} which will also be discussed in the context of Majorana representations in Secs.~\ref{sec:spin1/2} and \ref{sec:spin3}.

As a side remark concerning the notation, in this section it was convenient to introduce matrices $K^\mu_{mn}$ and vectors of operators such as $f_m$ with an index range $m,n\in\{-s,-s+1,\ldots,s\}$. For the discussion of Majorana representations in the following sections it turns out more natural to define indexed objects in the range $m,n\in\{1,\ldots,d\}$ where $d=2s+1$. As an adaption to this, from now on we will also use $K^\mu_{mn}$ and $f_m$ from this section with a shifted index range $m,n\in\{1,\ldots,d\}$, i.e., where the same entries appear at indices shifted by $s+1$.

\section{General construction of bilinear Majorana spin representations}\label{majoconstruction}
We continue by presenting a systematic and exhaustive formalism for constructing bilinear Majorana representations for spin operators, which can be used to describe spins of magnitudes $S\geq 1/2$. To this end, we define $d \in\mathbb{N}$ Majorana operators $c_1, \ldots, c_d$, which satisfy the following defining relations:
\begin{align}
	\{c_m, c_n\} &= 2\delta_{mn}, \\ \label{cliffordrelation}
	c_n^\dagger &= c_n,
\end{align}
with $m,n\in\{1,\ldots,d\}$. The most general ansatz for a bilinear Majorana representation of a spin operator $\bm S$ then has the form
\begin{align}
	S^\mu = \frac{1}{4}c_m \tilde{K}^\mu_{mn} \, c_n\equiv\frac{1}{4}C^T \tilde{K}^\mu \, C. \label{ansatz}
\end{align}
In the rightmost term we have introduced a shorthand matrix notation where $C=(c_1,\ldots,c_d)^T$ is a $d$-tuple Majorana operator. Crucially, $\tilde{K}^\mu$ is a skew-symmetric (i.e., $\tilde{K}^{\mu}_{mn}=-\tilde{K}^{\mu}_{nm}$) and Hermitian matrix, which requires all elements to be purely imaginary. Consequently, diagonal elements vanish, $\tilde{K}^{\mu}_{nn}=0$.

Enforcing the spin algebra of Eq.~(\ref{spinalgebra1}) one easily finds that the matrices $\tilde{K}^\mu$ have to fulfill the same type of condition upon matrix multiplication
\begin{align}
	[\tilde{K}^\mu,\tilde{K}^\nu] = \mathrm{i} \epsilon^{\mu \nu \rho} \tilde{K}^\rho, \label{spinalgebra2}
\end{align}
implying that $\tilde{\bm K}=(\tilde{K}^x,\tilde{K}^y,\tilde{K}^z)$, like ${\bm K}$ in Sec.~\ref{sec:complex_fermions}, are generators of a matrix representation of SU(2) with elements $e^{-\mathrm{i}\varphi {\bm n}\cdot\tilde{\bm K}}$. However, with $\tilde{K}^\mu$ purely imaginary, obviously these matrix representations are purely real. Consequently, the bilinear Majorana representations for spin operators are determined by the real representations of SU(2). While the complex irreducible representations of SU(2) from Sec.~\ref{sec:complex_fermions} are ubiquitous in physics, the real ones are rarely considered, but they have been classified mathematically \cite{itzkowitz_note_1991}: Up to unitary transformations, there exists exactly one real irreducible representation of SU(2) for each dimension $d$ that is either odd or divisible by four, while for dimensions divisible by two but not by four there only exist reducible representations.

Since (up to unitary transformations) the matrices $K^\mu$ in Eq.~(\ref{eq:matrixrepn}) already cover all generator matrices of SU(2) the remaining task to find $\tilde{K}^\mu$ is to identify a unitary transformation matrix $T$ which turns $K^\mu$ into purely imaginary matrices,
\begin{align}
	\tilde{K}^\mu = T^\dagger K^\mu T. \label{Ttrafo}
\end{align}
According to the aforementioned theorem, this way of constructing irreducible and imaginary matrices $\tilde{K}^\mu$ only works if their dimension $d$ is either odd or divisible by four. In the following subsection we start discussing the case of odd $d$. As will also be explained below, even though the matrices $K^\mu$ and $\tilde{K}^\mu$ are just related by a unitary transformation, the corresponding spin representations in Eqs.~(\ref{eq:complex_rep1}) and (\ref{ansatz}) realize different sets of spin spaces $\mathcal{M}(s)$.

Generally, one can also consider Majorana spin representations resulting from reducible matrices $\tilde{K}^\mu$. Decomposing these reducible matrices $\tilde{K}^\mu$ into irreducible blocks, each block itself corresponds to a spin representation which together form a product space of individual spins. Such representations have been employed in previous literature~\cite{wen99,barkeshli10,baez_numerical_2017,ma22} and are often very helpful, however, here we do not investigate this case further since the reducible representations can be straightforwardly constructed based on the irreducible ones. 

\subsection{Odd-dimensional representations} \label{sec:odd}
When $d$ is odd, we can assign an integer spin $s=(d-1)/2\in\mathbb{N}$ to the $d\times d$ matrices $K^\mu$ and $\tilde{K}^\mu$ via Eq.~(\ref{eq:matrixrepn}). In the following, we will describe a general procedure to construct the unitary transformation matrix $T$ [see Eq.~(\ref{Ttrafo})] for arbitrary $s\in\mathbb{N}$. We first write $T$ as a product $T=PQ$. Here, $P$ is a permutation, introduced for notational convenience, which rearranges the diagonal elements of $K^z$ from the consecutive order $s,s-1,\ldots,-s$ [see Eq.~(\ref{eq:iz})] to a pairwise and descending sequence of elements $(n,-n)$ where the single eigenvalue $n=0$ is put at the end,
\begin{align}
	P^TK^zP = \mathrm{diag}(s,-s,s-1,-s+1,\ldots,1,-1, 0) \label{eq:PKP}.
\end{align}
This way, $P^TK^zP$ has a $2\times2$ block structure where each block is proportional to a Pauli spin-$1/2$ $\sigma^z$ matrix (except of the last trivial $1\times1$ block). Next, $Q$ is defined as
\begin{align}
	Q = \left[\bigoplus_{n=1}^s \, \frac{\mathrm{e}^{\mathrm{i} \theta_n}}{\sqrt{2}}\begin{pmatrix}
	-\mathrm{i} & 1 \\
	\phantom{-}\mathrm{i} & 1 
\end{pmatrix}\right] \oplus1. \label{eq:Q}
\end{align}
Applying $Q$ to Eq.~(\ref{eq:PKP}), to get $\tilde{K}^z=T^\dagger K^z T$, transforms each $\sigma^z$ block in $P^TK^zP$ into a block proportional to $\sigma^y$ while the phase factors $\mathrm{e}^{\mathrm{i} \theta_n}$ drop out (the trivial $1\times1$ block remains unaffected). The matrix $\tilde{K}^z$ then reads as,
\begin{equation}
\tilde{K}^z=\left(\begin{array}{ccccc}
\left(\begin{array}{cc}0&\mathrm{i}s\\-\mathrm{i}s&0\end{array}\right)&&&&\\
&\ddots&&&\\
&&\left(\begin{array}{cc}0&2\mathrm{i}\\-2\mathrm{i}&0\end{array}\right)&&\\
&&&\left(\begin{array}{cc}0&\mathrm{i}\\-\mathrm{i}&0\end{array}\right)&\\
&&&&(0)
\end{array}\right).
\end{equation}
Most importantly, with these $\sigma^y$ blocks on the diagonal the resulting matrix $\tilde{K}^z$ is purely imaginary, as desired.
Indeed, the block structure of $\tilde{K}^z$ will turn out very useful below. The remaining task to also ensure that $\tilde{K}^x$ and $\tilde{K}^y$ are purely imaginary after the transformation with $Q$ is to properly adjust the phases $\theta_n$ (with $n\in\{1,\ldots,s\}$) which do not drop out in these two matrices. A simple choice which accomplishes this is given by $\theta_n = (s-n+1)\pi/2$, however, more possibilities can be identified but we refrain from discussing them here.

With the knowledge of a matrix $T$ and with Eqs.~(\ref{ansatz}) and (\ref{Ttrafo}) the problem of finding bilinear Majorana representations of spin operators from odd-dimensional matrices $K_\mu$ is formally solved at this point:
\begin{equation}\label{majorepn0}
 	S^\mu = \frac{1}{4} C^T T^\dagger  K^\mu  T  C.    
\end{equation}
The result is unique up to orthogonal O($d$) transformations of all $\tilde{K}^\mu$ with $\mu\in\{x,y,z\}$. Even though such additional transformations do not change the physical properties of the Majorana presentations, it will turn out useful to apply a subgroup of O($d$) transformations leading to a somewhat more general version Eq.~(\ref{majorepn0}). Particularly, we consider an additional rotation $R$ such that Eq.~(\ref{majorepn0}) is replaced by
\begin{align}
	S^\mu = \frac{1}{4} C^T R^T  T^\dagger  K^\mu  T  R \, C, \label{majorepn}
\end{align}
where $R$ is given by
\begin{align}
	R = \left[\bigoplus_{n=1}^s \Big(\mathrm{e}^{-\mathrm{i} \phi_n\sigma^y}\Big)\right] \oplus 1. \label{rotation}
\end{align}
The $2\times2$ blocks $\mathrm{e}^{-\mathrm{i} \phi_n\sigma^y}$ in $R$ are real 2-dimensional SO($2$) rotation matrices. Since $T^\dagger K^z T$ consists of $\sigma^y$ blocks, the additional transformation $R$ does not change $S^z$ in Eq.~(\ref{majorepn}) compared to Eq.~(\ref{majorepn0}) but leads to more general representations for $S^x$ and $S^y$ where $\phi_n\in[0,2\pi)$ are free parameters. Written out explicitly in terms of Majorana operators, Eq.~(\ref{majorepn}) takes the following final form

\begin{widetext}
\begin{align}
	S^x &= -\frac{\mathrm{i}}{\sqrt{2}} \big(A_s c_{2s-1} +  B_s c_{2s}\big) \, c_{2s+1} + \frac{\mathrm{i}}{2}\sum_{n=1}^{s-1} \Big[A_{n}\big( c_{2n}c_{2n+1}-c_{2n-1}c_{2n+2}\big) - B_{n}\big(c_{2n-1}c_{2n+1}+c_{2n}c_{2n+2}\big)\Big], \label{majorepnx} \\ 
	S^y&= -\frac{\mathrm{i}}{\sqrt{2}} \big(B_s c_{2s-1}-A_s c_{2s}\big) \, c_{2s+1} + \frac{\mathrm{i}}{2}\sum_{n=1}^{s-1} \Big[B_{n}\big(c_{2n}c_{2n+1}-c_{2n-1}c_{2n+2}\big) + A_{n}\big( c_{2n-1}c_{2n+1}+c_{2n}c_{2n+2}\big) \Big],\label{majorepny} \\ 
	S^z&= \frac{\mathrm{i}}{2} \sum_{n=1}^{s} \, (s-n+1) \, c_{2n-1}c_{2n}, \label{majorepnz}
\end{align}
\end{widetext}
where
\begin{align}
	A_n &=  \frac{1}{2} \mathrm{sin}(\phi_n-\phi_{n+1})\sqrt{n(2s-n+1)}, \notag\\
	B_n &= \frac{1}{2}\mathrm{cos}(\phi_n-\phi_{n+1}) \sqrt{n(2s-n+1)}, \label{eq:B}
\end{align}
with $n\in\{1,\ldots,s\}$ and $\phi_{s+1}=0$. Note that the square root expressions in Eq.~(\ref{eq:B}) are the ones from Eq.~(\ref{eq:ixiy}) under the index shift $m=s-n$. Choosing the angles $\phi_n$ such that either the $\sin$ or $\cos$ terms in Eq.~(\ref{eq:B}) vanish, one obtains Majorana representations with the minimal number of terms.  

From the discussion so far it is clear that our Majorana representations fulfill the SU(2) spin algebra and are therefore proper rewritings of spin operators. However, with an odd number of Majorana fermions the many particle Hilbert space has the formal dimension $2^{d/2}$, which is irrational. To be able to straightforwardly define an action of $S^\mu$ on an integer dimensional Hilbert space it is convenient to introduce an additional Majorana operator $c_{2s+2}$. For a lattice spin system, a physically motivated way of introducing this extra Majorana fermion is to borrow it from another site which means that $c_{2s+2}$ on site $i$ is associated with one of the operators $c_1,\ldots,c_{2s+1}$ on site $j$ and vice versa. This way, the Hilbert space for the pair of sites $(i,j)$ is described by an even number of Majorana fermions and has the integer dimension $2^d$. For spin-$1/2$ Majorana representations this procedure has previously been applied in Refs.~\cite{biswas_su2-invariant_2011,fu_majorana_2018}. Alternatively, the extra Majorana operator $c_{2s+2}$ can just be a ``dummy'' operator which has no further physical meaning (see e.g. Ref.~\cite{shnirman03}). The following discussions include $c_{2s+2}$ but we leave its origin unspecified.

We now discuss which spin magnitudes $S\in\mathcal{M}(s)$ are generated by the Majorana representation in Eqs.~(\ref{majorepnx})-(\ref{majorepnz}). To this end, we proceed similarly to Sec.~\ref{sec:complex_fermions}, i.e., we consider the set of eigenvalues $M^z$ of $S^z$. This set consists of subsets of consecutive numbers $S,S-1,\ldots,-S$ which indicate the realized magnitudes $S$. Crucially, we profit from the $2\times2$ block diagonal form of $\tilde{K}^z$, which means that $S^z$ decomposes into fermionic parity terms $\mathrm{i} c_{2n-1}c_{2n}$ with $n=1,\ldots,s+1$, see Eq.~(\ref{majorepnz}). These operators all commute with each other and have eigenvalues $\lambda_n = \pm1$. The corresponding eigenstates $|\lambda_1 ... \lambda_{s+1} \rangle$ with 
\begin{align}
	\mathrm{i} c_{2n-1}c_{2n} \, |\lambda_1 ... \lambda_{s+1} \rangle = \lambda_n \, |\lambda_1 ... \lambda_{s+1} \rangle, \label{eq:defeigenvalues}
\end{align}
are a basis in the $2^{s+1}=2^{(d+1)/2}$ dimensional Hilbert space. The set of eigenvalues of $M^z$ is then given by
\begin{equation}\label{eq:mzmajo}
 M^z\in \Bigg\{\sum_{m=0}^{s} \frac{m}{2} \lambda_m \Bigg\}\Bigg|_{\{\lambda_m=\pm1\}}.
\end{equation}
[Note that this set correctly captures all quantum numbers of $M^z$ but to lighten the notation we do not adhere to a precise matching between the contributions from each $\lambda_m$ in Eqs.~(\ref{majorepnz}) and (\ref{eq:mzmajo})]. It is instructive to compare this set with the corresponding one from a complex fermionic representation in Eq.~(\ref{eq:mz}) for the same value of $s$. Rewriting the occupation numbers of complex fermions $N_m$ in terms of a quantum number $\lambda_m$ with eigenvalues $\pm1$ via $\lambda_m=2N_m-1=\pm1$ one finds an analogous expression,
\begin{equation}
 M^z\in \Bigg\{\sum_{m=-s}^{s} \frac{m}{2} \lambda_m \Bigg\}\Bigg|_{\{\lambda_m=\pm1\}}.
\end{equation}
Here, only the fact that $s$ can also be half-integer and the range of the sum differ from Eq.~(\ref{eq:mzmajo}). This indicates a more complicated structure of spin subspaces in the complex fermionic case than in the Majorana case.

Coming back to the set of quantum numbers $M^z$ for Majorana fermions in Eq.~(\ref{eq:mzmajo}) the vanishing of the term with $m=0$ implies that all eigenvalues of $M^z$ have a degeneracy, which is a multiple of two. Equivalently, the Hilbert space splits into two identical copies, which is signalled by the operator $\mathrm{i}c_{2s+1}c_{2s+2}=\pm1$ not explicitly appearing in Eq.~(\ref{majorepnz}). These two identical subspaces can be distinguished by the eigenvalues of the operator
\begin{equation}
D=(\mathrm{i}c_1c_2)(\mathrm{i}c_{3}c_{4})\cdots(\mathrm{i}c_{2s+1}c_{2s+2})=\pm1,\label{eq:d} 
\end{equation}
which commutes with all spin components in Eqs.~(\ref{majorepnx})-(\ref{majorepnz}), $[S^\mu,D]=0$ and which is a generalization of the operator $D$ known from Kitaev's Majorana representation for spin-1/2 (see Ref.~\cite{kitaev_anyons_2006}).

Setting all eigenvalues $\lambda_n=+1$ the largest realized spin magnitude $S_\text{max}$ in a Majorana representation for a given $s$ is found to be
\begin{equation}\label{eq:smax}
    S_\text{max}=\frac{1}{2}\sum_{m=1}^s m=\frac{s(s+1)}{4}=\frac{d^2-1}{16},
\end{equation}
and, apart from the two copies resulting from $D=\pm1$, there is no other subspace with this spin magnitude.

Since auxiliary particle representations for spin operators are primarily meant to realize a single spin subspace, it is an important question whether there are values $s$ for which no other subspaces with $S<S_\text{max}$ exist. Considering the terms in Eq.~(\ref{eq:mzmajo}),
\begin{equation}
    \pm\frac{1}{2}\pm1\pm\frac{3}{2}\pm2\pm\ldots,\label{eq:sum}
\end{equation}
whenever two different combinations of signs yield the same contribution there must at least be one additional spin space with $S<S_\text{max}$. It is easy to see that this first happens when $\pm3/2$ is part of the sum, since then, the same contribution $1/2+1-3/2=-1/2-1+3/2=0$ appears for two different combinations of signs. This already excludes the existence of a single sequence $-S,-S+1,\ldots,S$ in the set of Eq.~(\ref{eq:mzmajo}) for all $s>2$ and additional subspaces are necessarily generated. It follows that (up to the two redundant sectors from $D=\pm1$) only the cases $s=1$ and $s=2$ realize ``pure'' Majorana spin representations with a single spin amplitude, namely $S=1/2$ and $S=3/2$, respectively, which will be further discussed in Sec.~\ref{sec:examples}. We list all realized spin spaces of these Majorana representations and the complex fermionic representations for $s\leq5$ in Table~\ref{tab5} in the Appendix which clearly demonstrates a more complex structure of subspaces in the complex fermionic case than in the Majorana case.

\subsection{Representations with dimensions $d$ divisible by four}\label{sec:divisible_by_four}
When the dimension $d$ of $\tilde{K}^\mu$ is divisible by four, one can in principle proceed in the same way as in Eq.~(\ref{Ttrafo}), i.e., by starting with the complex $d\times d$ matrix $K^\mu$ and performing a unitary transformation with a matrix $T$ to yield a purely imaginary matrix $\tilde{K}^\mu$. However, as mentioned in Ref.~\cite{itzkowitz_note_1991} a simpler method is possible, which starts with the $d/2$-dimensional complex matrix $K^\mu$ from Eq.~(\ref{eq:matrixrepn}) and transforms it to a $d$-dimensional matrix $\tilde{K}^\mu$ in a procedure called {\it realification}. More precisely, the realification $\mathcal{R}(H)$ of a $d/2$-dimensional matrix $H$ substitutes each complex element $H_{mn}=H_{mn}'+iH_{mn}''$ with $H_{mn}',H_{mn}''\in\mathbb{R}$ by a $2\times2$ matrix according to
\begin{align}
&\mathcal{R}:\mathbb{C}^{\frac{d}{2}\times\frac{d}{2}}\rightarrow\mathbb{R}^{d\times d}, \notag\\&H_{mn}\mapsto H_{mn}'\left(\begin{array}{cc}1&0\\0&1\end{array}\right)+H_{mn}''\left(\begin{array}{cc}0&-1\\1&0\end{array}\right),\label{eq:def_realification}
\end{align}
hence doubling its dimension. The desired $d$-dimensional and purely imaginary matrices $\tilde{K}^\mu$ which define the Majorana representation in Eq.~(\ref{ansatz}) then follow from the $d/2$-dimensional complex matrix $K^\mu$ via
\begin{equation}
\tilde{K}^\mu=\mathrm{i}\mathcal{R}(-\mathrm{i}K^\mu). \label{eq:realification}
\end{equation}
Crucially, the procedure of realification preserves the spin commutation algebra.

However, it is straightforward to see that the $d$ dimensional Majorana representations for spin operators constructed this way are not new. Indeed, they are equivalent to the $d/2$-dimensional complex fermionic representations defined in Eq.~(\ref{eq:complex_rep1}) when the matrix $K^\mu$ in Eq.~(\ref{eq:complex_rep1}) is the same as the $d/2$ dimensional matrix $K^\mu$ that is used in Eq.~(\ref{eq:realification}). This is due to the fact that the procedure of realification in Eq.~(\ref{eq:realification}) is identical to the standard mapping between complex fermions and Majorana fermions via 
\begin{equation}
f_n=\frac{1}{2}(c_{2n-1}+\mathrm{i}c_{2n}),\quad
f_n^\dagger=\frac{1}{2}(c_{2n-1}-\mathrm{i}c_{2n}), \label{eq:MajoranaRewriting2}
\end{equation}
where $n\in\{1,\ldots,d/2\}$ and the fermionic particle number operator becomes $N_n=f_n^\dagger f_n= (1+\mathrm{i}c_{2n-1}c_{2n})/2=(1+\lambda_n)/2$. Particularly, this mapping does not change the generated spin subspaces $S$.

As discussed in Sec.~\ref{sec:complex_fermions}, the complex $d/2$ dimensional fermionic representations that correspond to these $d$ dimensional Majorana representations realize a half-odd integer spin $S=(d/2-1)/2$ in the subspace of single fermionic occupancy. In the case $d=4$ this is the Abrikosov representation for spin-$1/2$~\cite{abrikosov_electron_1965} and the associated $4$-dimensional Majorana representation has been coined the SO(4) chiral representation in Ref.~\cite{fu_majorana_2018}. 

This concludes the classification of all bilinear Majorana representations of spin operators, which follow from irreducible spin matrices. While for $d\,\textrm{mod }4 = 0$, as discussed in this subsection, no new representations are found and hence, this case will not be considered any further, the case of odd $d$ from the previous subsection is worth studying in more detail, which will be done in Sec.~\ref{sec:examples}. Importantly, the representations from odd $d$ cannot be brought into the complex fermionic form of Eq.~(\ref{eq:complex_rep1}) via the mapping in Eq.~(\ref{eq:MajoranaRewriting2}), indicating that they belong to a separate class of Majorana representations. Since complex fermionic spin representations never realize a single spin subspace $S$, it also follows that spin-$1/2$ and spin-$3/2$ operators are the only ones which allow for a  Majorana rewriting that does not come along with additional spin subspaces (except of a redundancy from the double degeneracy due to $D=\pm1$, see discussion in Sec.~\ref{sec:odd}).

\subsection{Representations with dimensions $d$ divisible by two but not by four}\label{sec:divisible_by_two}
It is worth briefly commenting on the case $d\,\textrm{mod }4 = 2$, even though the sought-after real irreducible matrix representations of SU(2) do not exist for such dimensions~\cite{itzkowitz_note_1991}. However, the realification procedure of $d/2$ dimensional matrices $K^\mu$ via Eq.~(\ref{eq:realification}) generates $d$ dimensional real matrices $\tilde{K}^\mu$ that fulfill the commutator algebra in Eq.~(\ref{spinalgebra2}). Obviously, these matrices $\tilde{K}^\mu$ have to be reducible and one, indeed, finds them to decompose into two identical $d/2$ dimensional real irreducible matrices $\tilde{k}^\mu$, i.e., $\tilde{K}^\mu=\tilde{k}^\mu\oplus\tilde{k}^\mu$~\cite{itzkowitz_note_1991}. Since $d/2$ is odd, the $d/2$ dimensional Majorana representations that follow when inserting $\tilde{k}^\mu$ into Eq.~(\ref{ansatz}) must already be covered by the odd dimensional representations discussed in Sec.~\ref{sec:odd}. This means that (up to orthogonal transformations) the Majorana representation from $\tilde{k}^\mu$ is identical to the one that is obtained from a $d/2$ dimensional matrix $\tilde{K}^\mu$ via the transformation procedure described in Eq.~(\ref{Ttrafo}).

This provides an alternative construction of the odd dimensional Majorana representations from Sec.~\ref{sec:odd}: One starts with a complex fermionic representation in Eq.~(\ref{eq:complex_rep1}), where $K^\mu$ is odd ($d/2$) dimensional. This representation is then rewritten in terms of $d$ Majorana fermions using Eq.~(\ref{eq:MajoranaRewriting2}). For certain orthogonal transformations of the Majorana vector $C=(c_1,\ldots, c_d)^T$ this representation decomposes into two identical odd-dimensional representations, each of which corresponds to those found in Sec.~\ref{sec:odd}. Note that this reducibility into two identical representations only becomes possible {\it after} rewriting the complex fermions as Majorana fermions via Eq.~(\ref{eq:MajoranaRewriting2}) and cannot be performed through a unitary transformation of the vector of complex fermions $(f_1,\ldots,f_{d/2})$. The fact that the odd dimensional complex fermionic representations in Eq.~(\ref{eq:complex_rep1}), which realize an integer spin in the singly occupied fermionic subspace, can be understood as being composed of two identical Majorana representations is still indicated by their spectra of realized spin amplitudes $S$, see Table~\ref{tab5} in the Appendix. For example, up to two-fold degeneracies, the spin spaces $3\oplus2\oplus1\oplus0$ of a complex fermionic representation with $s=2$ can be viewed as resulting from the $S=3/2$ Majorana spin space for the same value $s=2$ when forming the combination $3/2\otimes 3/2=3\oplus2\oplus1\oplus0$, and similarly for all other integer values $s$. Furthermore, subspaces $S$ of complex fermionic representations with integer $s$ are always doubly degenerate, a property which is directly inherited from the underlying odd dimensional Majorana representations where $D=\pm1$ generates a double degeneracy.

\section{Examples for explicit Majorana spin representations} \label{sec:examples}
Having classified all bilinear Majorana spin representations, we will now study their properties in more detail by discussing explicit examples with small dimensions $d$ and associating them with known representations, if possible. As explained in Sec.~\ref{sec:divisible_by_four}, the case where the dimension $d=2s+1$ is divisible by four will not be considered further, except of a brief discussion of the SO(4) representation for spin-$1/2$ in Sec.~\ref{sec:spin1/2}. Concretely, in the following Secs.~\ref{sec:spin1/2}-\ref{sec:spin3} we will focus on the representations in Eqs.~(\ref{majorepnx})-(\ref{majorepnz}) for $s=1,2,3$, respectively, and in Sec.~\ref{sec:higherspin} we will briefly discuss general properties for arbitrary $s\in\mathbb{N}$.

\subsection{Spin-$1/2$ realized for $s=1$}\label{sec:spin1/2}
For $s=1$ (i.e., $d=3$), the construction in Sec.~\ref{sec:odd} provides a Majorana representation for $S=1/2$ spin operators in terms of three Majorana fermions $c_1$, $c_2$, $c_3$ and possibly a fourth operator $c_4$ to define a four dimensional Hilbert space they act on. Fixing the only leftover parameter $\phi_1 = \pi/2$ in Eq.~(\ref{eq:B}) and renaming $c_1\equiv b^y$, $c_2\equiv b^x$, $c_3\equiv b^z$, $c_4\equiv c$ in Eqs.~(\ref{majorepnx})-(\ref{majorepnz}) one finds that this is the known SO(3) Majorana representation for spin-$1/2$ operators~\cite{martin59,tsvelik92},
\begin{align}
	S^\mu = -\frac{\mathrm{i}}{4} \epsilon^{\mu \nu \rho} b^\nu b^\rho. \label{so3repn}
\end{align}
Even though this representation is well-known we find it useful to briefly discuss it here from our perspective and relate it to other known fermionic spin-$1/2$ representations, see also Ref.~\cite{fu_majorana_2018}. As already discussed in Sec.~\ref{sec:odd}, this representation does not produce any additional spin spaces with $S\neq1/2$, but comes along with a redundancy in the form of a two-fold degeneracy. More explicitly, when $S^\mu$ from Eq.~(\ref{so3repn}) is written as a matrix $\underline{S}^\mu$ in the basis
\begin{equation}\label{eq:basis}
|S^z,D\rangle\in\{|1/2,1\rangle,|-1/2,1\rangle,|1/2,-1\rangle,|-1/2,-1\rangle\},
\end{equation}
where $S^z=-\mathrm{i}b^x b^y/2$ and $D=b^x b^y b^z c$ [see Ref.~\cite{kitaev_anyons_2006} and Eq.~(\ref{eq:d})] one obtains two identical $2\times2$ blocks
\begin{equation}\label{mbmM1/2}
 \underline{S}^\mu=\frac{\sigma^\mu}{2}\oplus\frac{\sigma^\mu}{2}\,  
\end{equation}
where $\sigma^\mu$ are the standard Pauli matrices.

One can now construct Kitaev's spin-$1/2$ representation~\cite{kitaev_anyons_2006}, denoted $S^\mu_\text{K}$, by applying $D$ to Eq.~(\ref{so3repn}),
\begin{align}
	S_\text{K}^\mu = D S^\mu = \frac{\mathrm{i}}{2} b^\mu c.\label{eq:kitaev}
\end{align}
In a matrix representation using the basis from Eq.~(\ref{eq:basis}), the operators $S_\text{K}^\mu$ have the form
\begin{equation}\label{mbmK1/2}
 \underline{S}_\text{K}^\mu=\frac{\sigma^\mu}{2}\oplus\left(-\frac{\sigma^\mu}{2}\right),  
\end{equation}
where, compared to Eq.~(\ref{mbmM1/2}), the second $2\times2$ block has acquired a minus sign for all spin components. The action of $D$ in the subspace with $D=-1$ can be interpreted as time reversal $\mathcal{T}$ which transforms a spin $\bm{S}$ as $\mathcal{T}:\bm{S}\longrightarrow -\bm{S}$. Hence, the Kitaev representation is special in the sense that it yields two copies of spin-$1/2$ degrees of freedom, where one copy corresponds to a time reversed spin-$1/2$ that fulfills a slightly modified version of the spin algebra in Eq.~(\ref{spinalgebra1}), i.e., $[S^\mu, S^\nu] = -\mathrm{i} \epsilon^{\mu \nu \rho} S^\rho$. Since in spin lattice models (e.g. the Kitaev honeycomb model~\cite{kitaev_anyons_2006}), mixing between the two subspaces should be avoided, contributions with $D=-1$ have to be projected out, which can formally be achieved with the projector $P=(1+D)/2$.

The spin-$1/2$ Abrikosov representation and the equivalent SO(4) Majorana representation~\cite{chen12}, labelled $S^\mu_\text{A}$, now straightforwardly follow via the (normalized) sum of Eqs.~(\ref{so3repn}) and (\ref{eq:kitaev}),
\begin{equation}\label{eq:abrikosov}
    S^\mu_\text{A}=\frac{1}{2}(S^\mu+S^\mu_\text{K})=PS^\mu=PS^\mu P=PS^\mu_\text{K} P.
\end{equation}
The last two expressions in Eq.~(\ref{eq:abrikosov}) demonstrate that $S^\mu_\text{A}$ can be interpreted as a representation where the projection onto the $D=1$ subspace has been carried out on the operator level, while states with $D=-1$ are still part of the Hilbert space. This also manifests in the matrix representation for $S_\text{A}^\mu$, corresponding to the sum of Eqs.~(\ref{mbmM1/2}) and (\ref{mbmK1/2}),
\begin{equation}\label{mbmA1/2}
 \underline{S}_\text{A}^\mu=\frac{\sigma^\mu}{2}\oplus 0\oplus 0,  
\end{equation}
which features a two-dimensional subspace where $S^\mu_\text{A}$ vanishes. In other words, the $D=-1$ subspace hosts two trivial $S=0$ singlet states.

The precise correspondence between Eq.~(\ref{eq:abrikosov}), formulated in terms of Majorana operators $b^\mu$, $c$, and the usual complex fermionic version of Abrikosov's representation [see also Eq.~(\ref{eq:complex_rep1})],
\begin{equation}\label{eq:abrikosov2}
    S^\mu_\text{A}=\frac{1}{2} f^\dagger_n \sigma_{nm}^\mu f_m,
\end{equation}
with two fermion operators $f_\uparrow$ and $f_\downarrow$, is established by
\begin{equation}\label{eq:fc_trafo}
f_\uparrow=\frac{1}{2}(b^x-\mathrm{i}b^y),\quad
f_\downarrow=\frac{1}{2}(-b^z+\mathrm{i}c), 
\end{equation}
see also Refs.~\cite{burnell11,chen12,fu_majorana_2018}. The $S=1/2$ subspace with $D=1$ is then characterized by a fermionic particle number $N_f=f^\dagger_\uparrow f_\uparrow+f^\dagger_\downarrow f_\downarrow=1$ while the two spin-0 subspaces with $D=-1$ correspond to particle numbers $N_f=0$ and $N_f=2$. The spin spaces realized for the different spin-$1/2$ representations discussed here are summarized in Table~\ref{tab1}.
\begin{table}[t]
	\footnotesize
	\renewcommand\arraystretch{1.8}
	\setlength\tabcolsep{0.15cm}
	\begin{tabular}{|r|l|}
		\hline
		\multicolumn{2}{|c|}{\textbf{ (a) Spin-1/2}} \\
		\hline
		\hline
		Representation & Spin spaces  \\
		\hline
		Majorana ($s=1$) & $\frac{1}{2}\oplus\frac{1}{2}$ \\
		Kitaev ($s=1$)& $\frac{1}{2} \oplus \frac{1}{2}^\mathcal{T}$  \\
		Complex ($s=\frac{1}{2}$) & $0 \oplus \frac{1}{2} \oplus 0$  \\
		\hline
		\hline
		\multicolumn{2}{|c|}{\textbf{(b) Spin-3/2}} \\
		\hline
		\hline
		Representation & Spin spaces \\
		\hline
		Majorana ($s=2$) & $\frac{3}{2} \oplus \frac{3}{2}$  \\
		Kitaev ($s=2$) & $\frac{3}{2} \oplus \frac{3}{2}^\mathcal{T}$ \\
		Complex ($s=\frac{3}{2}$) & $0 \oplus \frac{3}{2} \oplus (2 \oplus 0) \oplus \frac{3}{2} \oplus 0  $ \\
		\hline
		\hline
		\multicolumn{2}{|c|}{\textbf{(c) Spin-3}} \\
		\hline
		\hline
		Representation & Spin spaces \\
		\hline
		Majorana ($s=3$)& $3 \oplus 0 \oplus \mathrm{copies}$  \\
		Kitaev ($s=3$) & $3 \oplus 0 \oplus \mathrm{copies}^\mathcal{T}$ \\
		Complex ($s=3$) & $0 \oplus 3 \oplus (5 \oplus 3 \oplus 1) \oplus (6 \oplus 4 \oplus 3 \oplus 2 \oplus 0)$  \\
		& $\oplus$ $\mathrm{copies}$\\
		\hline
	\end{tabular}
	\caption{Classification of possible spin representations realizing subspaces with (a) spin-1/2, (b) spin-3/2 and (c) spin-3, together with other spin spaces, as listed. The Majorana representations correspond to the ones of Eqs.~(\ref{majorepnx})-(\ref{majorepnz}) for a given $s$ while the Kitaev representations result from the former upon application of $D$ [see Eq.~(\ref{eq:d})]. The label ``$\mathcal{T}$'' indicates a time-reversed spin operator. The representations named ``complex'' refer to the complex fermionic representations from Eq.~(\ref{eq:complex_rep1}) where $s$ is chosen such that it provides spin-$1/2$, spin-$3/2$, or spin-$3$ subspaces, respectively, in the singly occupied sector with $N_f=1$. The complex fermionic spin-$1/2$ representation is the Abrikosov representation where the subspaces correspond to particle numbers $N_f = 0,1,2$ from left to right. Spin spaces in brackets have the same particle number $N_f$ (e.g. the two spaces in brackets for the complex fermionic spin-$3/2$ representation have $N_f=2$), and ``copies'' refers to an exact doubling of all preceding sectors.}
	\label{tab1}
\end{table}

This discussion shows that for $s=1$ the complex fermionic representation in Eq.~(\ref{eq:complex_rep1}) and the Majorana representation in Eqs.~(\ref{majorepnx})-(\ref{majorepnz}) are simply related via a projection $P$ [see Eq.~(\ref{eq:abrikosov})] and the transformation in Eq.~(\ref{eq:fc_trafo}). It should be emphasized, however, that this type of projective relation between Eq.~(\ref{eq:complex_rep1}) and Eqs.~(\ref{majorepnx})-(\ref{majorepnz}) does not hold for larger $s>1$, since $PS^\mu$ for higher spins is no longer bilinear in the Majorana fermions. A more general relation between Eq.~(\ref{eq:complex_rep1}) and Eqs.~(\ref{majorepnx})-(\ref{majorepnz}) has been discussed in Sec.~\ref{sec:divisible_by_two}.

An elegant way of projecting out the spin-0 degrees of freedom in the Abrikosov representation also on the level of states in the Hilbert space is the Popov-Fedotov approach~\cite{popov88,prokofiev11,doi:10.1142/S0217979206033310}. Since we will discuss a generalization of this method in Sec.~\ref{sec:spin3} below, we briefly introduce it here. Let $H_f$ be a general spin-$1/2$ Hamiltonian for an arbitrary number of spins (e.g. defined on the sites of a lattice) and arbitrary spin interactions between them. Furthermore, all spin operators occurring in $H_f$ are assumed to be expressed in terms of Abrikosov fermions via Eq.~(\ref{eq:abrikosov2}). Within the Popov-Fedotov method $H_f$ is replaced by
\begin{equation}\label{eq:hpf}
    H_f \longrightarrow H'_f = H_f+H_\text{PF}
\end{equation}
where the additional term $H_\text{PF}$ is given by
\begin{equation}
    H_\text{PF}=\frac{\mathrm{i}\pi}{2\beta}N_f.
\end{equation}
Here, $N_f$ is the total fermion number just for one (arbitrarily selected) spin. In order to project out the spin-0 subspaces of all spins, terms $H_\text{PF}$ also have to be added for the other spins. However, for our purpose of demonstrating how the Popov-Fedotov method generally works, it is sufficient here to consider only a single spin for which the projection is carried out.

We evaluate the partition function $Z=\text{Tr }e^{-\beta H'_f}$ for the considered spin:
\begin{equation}\label{eq:pf}
Z\propto\bigg(\sum_{N_f=1}+\sum_{N_f=0,2}\bigg)e^{-\beta E_f}e^{-\frac{\mathrm{i}\pi}{2}N_f}.
\end{equation}
Here, the first sum goes over the two spin-$1/2$ states with $N_f=1$ while the second sum is over the spin-0 subspaces. Furthermore, $E_f$ are the eigenenergies of $H_f$. In the first sum, the term $e^{-\beta H_{\text{PF}}}$ only leads to an irrelevant factor $\mathrm{i}$. Crucially, in the second sum the two spin-0 subspaces with $N_f=0$ and $N_f=2$ lead to the same contribution from $e^{-\beta E_f}$. Consequently, the second sum in Eq.~(\ref{eq:pf}) can be put in front of $e^{-\beta H_\text{PF}}$ such that the mutual cancellation of the contributions from the two spin-0 subspaces becomes obvious,
\begin{equation}
    \sum_{N_f=0,2}e^{\frac{\mathrm{i}\pi}{2}N_f}=1-1=0,
\end{equation}
and the projection is carried out exactly.

\subsection{Spin-$3/2$ realized for $s=2$}\label{sec:spin3/2}
We continue discussing the next higher-dimensional Majorana representation for which $s=2$. According to Eq.~(\ref{eq:smax}) and the discussion below Eq.~(\ref{eq:sum}) it realizes a ``pure'' spin-$3/2$ operator without any subspaces of smaller spin magnitudes. Due to $d=2s+1=5$, the representation is based on five Majorana operators $c_1,\ldots,c_5$ and, as before, an extra operator $c_6$ may be introduced to define an eight dimensional Hilbert space. Choosing the free parameters in Eq.~(\ref{eq:B}) as  $\phi_1=0$ and $\phi_2=-\pi/2$ and renaming the Majorana operators as $c_1\equiv \eta^z$, $c_2\equiv\theta^z$, $c_3\equiv\eta^x$, $c_4\equiv\eta^y$, $c_5\equiv\theta^x$ and $c_6\equiv\theta^y$, Eqs.~(\ref{majorepnx})-(\ref{majorepnz}) reproduce a spin-$3/2$ representation that has been used in a number of previous works~\cite{wang_z_2009,yao09,yao11,chua11,natori_chiral_2016,natori_dynamics_2017,natori18,carvalho18,de_farias_quadrupolar_2020,natori20,jin_unveiling_2022,PhysRevB.83.060407,PhysRevX.12.041029},
\begin{align}
	S^x &= \frac{\mathrm{i}}{2} \Big[\eta^y\eta^z - \eta^x \big(\theta^z - \sqrt{3}\theta^x\big)\Big],\notag \\
	S^y &= \frac{\mathrm{i}}{2} \Big[\eta^z\eta^x - \eta^y \big(\theta^z + \sqrt{3}\theta^x\big)\Big],\notag \\
	S^z &= \frac{\mathrm{i}}{2} \Big(\eta^x\eta^y + 2\eta^z \theta^z\Big),\label{eq:spin3/2}
\end{align}
where it often appears in the context of spin-orbit coupled~\cite{wang_z_2009,yao11,chua11,natori_chiral_2016,natori_dynamics_2017,natori18,carvalho18,de_farias_quadrupolar_2020,natori20} and SU(4) symmetric~\cite{wang_z_2009,natori18} models. In contrast, here we identify the same spin-3/2 representation only on the basis of the spins' symmetry group SU(2).

The eigenvalues $\pm1$ of the operator $D$ in Eq.~(\ref{eq:d}) distinguish between two identical spin-$3/2$ subspaces of the eight dimensional Hilbert space. In analogy to Eq.~(\ref{eq:kitaev}) one can formulate a ``Kitaev-type'' spin-$3/2$ representation as $S^\mu_K = DS^\mu$, where the $D=-1$ subspace hosts a time-reversed spin-$3/2$. However, this representation is no longer bilinear in the Majorana fermions, which holds true for all $s>1$. One can also construct a Majorana spin-$3/2$ representation in the subspace with $D=1$ by applying the operator $D$ to only parts of the terms in Eq.~(\ref{eq:spin3/2}) as has been done in Ref.~\cite{jin_unveiling_2022}, yielding a combination of bilinear and quartic terms. Table~\ref{tab1} lists the realized spin spaces for Majorana and complex fermionic representations [see Eq.~(\ref{eq:complex_rep1})] for $S=3/2$. Particularly, the complex fermionic representation that realizes a spin $S=3/2$ in the singly occupied $N_f=1$ subspace (which requires $s=3/2$) is significantly more complicated than the Majorana representations and also contains subspaces with $S>3/2$, demonstrating an advantage of Majorana representations.

\subsection{Spin-$3$ realized for $s=3$}\label{sec:spin3}
Setting $s=3$ (i.e. $d=7$) in Eqs.~(\ref{majorepnx})-(\ref{majorepnz}), one obtains the lowest dimensional Majorana representation that realizes subspaces with different spin amplitudes $S$. Listing all the magnetic quantum numbers according to Eq.~(\ref{eq:mzmajo}) one easily identifies a $S=3$ and a $S=0$ subspace, both doubly degenerate, which altogether form a 16 dimensional Hilbert space. For completeness, we provide here an explicit representation for $\phi_1=\phi_2=\phi_3=0$ [see Eq.~(\ref{eq:B})] in terms of seven Majorana operators $c_1,\ldots,c_7$,
\begin{align}
	S^x &= \1-\frac{\mathrm{i}}{2} \2\left[\1\sqrt{\frac{3}{2}}(c_1c_3+c_2c_4)\1+\1\sqrt{\frac{5}{2}}(c_3c_5+c_4c_6)\1+\1\sqrt{6}c_6c_7\1\right]\3,\notag \\
	S^y &= \1-\frac{\mathrm{i}}{2} \2\left[\1\sqrt{\frac{3}{2}}(c_1c_4-c_2c_3)\1+\1\sqrt{\frac{5}{2}}(c_3c_6-c_4c_5)\1+\1\sqrt{6}c_5c_7\1\right]\3,\notag \\
	S^z &= \1\frac{\mathrm{i}}{2} \left(3c_1c_2+2c_3c_4+c_5c_6\right).\label{eq:spin3}
\end{align}
As before, one may introduce an additional Majorana operator $c_8$ to define the fermion parity operator $D=c_1 c_2 \cdots c_8$, as already shown in Eq.~(\ref{eq:d}). Then $DS^\mu$ yields a non-bilinear Kitaev-type representation where one of the two $S=3$ subspaces is subject to time-reversal. An intricate and again non-bilinear representation is obtained by $(1+D)S^\mu/2$ which, apart from one $S=3$ subspace exhibits nine trivial $S=0$ singlet subspaces. Table~\ref{tab1} summarizes possible realizations of $S=3$ spin representations.

The Majorana spin representation in Eq.~(\ref{eq:spin3}) is an instructive example to discuss possible methods for projecting out unwanted spin subspaces (in this case two $S=0$ spaces) and to highlight difficulties that arise when trying to do so. Particularly, we consider a generalization of the Popov-Fedotov method which has been discussed in the context of the spin-$1/2$ Abrikosov representation in Sec.~\ref{sec:spin1/2}.

A first difficulty arises because, unlike for the Abrikosov representation, where the particle number $N_f$ is used to distinguish between the different spin spaces, for Majorana spin representations no bilinear Majorana operator exists whose eigenvalues characterize the individual spin spaces. This is because, if such an operator $G$ of the general form
\begin{equation}
    G=c_m g_{mn} c_n\label{eq:general_bilinear}
\end{equation}
existed (where $g$ is an imaginary, skew-symmetric $d\times d$ matrix), it would have to commute with all spin components $S^\mu$ in Majorana representation, $[S^\mu,G]=0$. This implies that the matrix $g_{mn}$ would have to commute with $\tilde{K}^\mu_{mn}$ [see Eq.~(\ref{ansatz})] upon matrix multiplication, $[\tilde{K}^\mu,g]=0$. Since $\tilde{K}^\mu$ are irreducible matrix representations for the generators of SU(2) the only matrix which commutes with all components of $\tilde{K}^\mu$ is the Casimir matrix $(\tilde{K}^x)^2+(\tilde{K}^y)^2+(\tilde{K}^z)^2=s(s+1)\mathbb{I}_{2s+1}$. This means that $g$ must be proportional to the identity matrix (which is also a consequence of Schur's lemma) in which case  Eq.~(\ref{eq:general_bilinear}) becomes a trivial constant, proving our claim that eigenvalues of operators that are bilinear in Majorana fermions cannot distinguish between different spin spaces. Note that this is in contrast to complex fermions where the identity matrix $\delta_{mn}$ in the bilinear expression $f_m^\dagger \delta_{mn} f_n$ still yields a non-trivial operator, namely the total fermion number.

This discussion shows that more complicated operators beyond bilinear ones are required to formulate a generalization of the Popov-Fedotov method for the present $s=3$ case. One possible set of operators to distinguish between the spin spaces $3\oplus3\oplus0\oplus0$ for $s=3$ is given by $D$ [see Eq.~(\ref{eq:d})] and $\bm{S}^2$ which assume eigenvalues
\begin{equation}\label{eq:subspaces}
(\bm{S}^2,D)=(12,1),(12,-1),(0,1),(0,-1)
\end{equation}
in these spaces, respectively. Note that $\bm{S}^2$ contains terms quartic in the Majorana fermions.

Based on the operators $\bm{S}^2$ and $D$ a Popov-Fedotov-like projection scheme can be formulated, e.g. by choosing $H_\text{PF}$ [see Eq.~(\ref{eq:hpf})] as
\begin{equation}
    H_\text{PF}=\frac{\mathrm{i}\pi}{24\beta}(\bm{S}^2-12)D.
\end{equation}
An analogous expression for the partition function $Z=\text{Tr }e^{-\beta(H_f+H_\text{PF})}$ as in Eq.~(\ref{eq:pf}) yields
\begin{equation}\label{eq:pf2}
Z\propto\bigg(\sum_{\bm{S}^2=12,D=\pm1}+\sum_{\bm{S}^2=0,D=\pm1}\bigg)e^{-\beta E_f}e^{\frac{\mathrm{i}\pi}{24}(12-\bm{S}^2)D },
\end{equation}
where the first (second) sum goes over all $S=3$ ($S=0$) states. In the first sum the term $e^{-\beta H_\text{PF}}$ always yields the identity, however, in the second sum
\begin{equation}
   \propto\sum_{D=\pm1}e^{\frac{\mathrm{i}\pi}{2}D}=\mathrm{i}-\mathrm{i}=0
\end{equation}
the desired mutual cancellation of contributions from the two $S=0$ subspaces takes place such that only contributions from the $S=3$ spin spaces remain.

While this demonstrates the general existence of a Popov-Fedotov-like scheme for Majorana spin representations (which can also be extended to representations with $s>3$) it also shows that such projection methods will likely not be useful for actual calculations since $H_\text{PF}$ contains complicated combinations of operators with products involving four to eight Majorana fermions. On the positive side, however, if one aims to implement spin-3 operators through the representation in Eq.~(\ref{eq:spin3}), the extra spin-0 states can be considered the mildest possible addition of ``unphysical'' states. Furthermore, because of the considerable difference in spin magnitude of the two subspaces, which can be expected to also translate into an energetic difference, one might wonder how much impact the additional spin sectors actually have on the low energy physics of a spin system. This question will be further discussed in Sec.~\ref{sec:states}.

\begin{figure*}
	\begin{center}
		\includegraphics[width=0.7\textwidth]{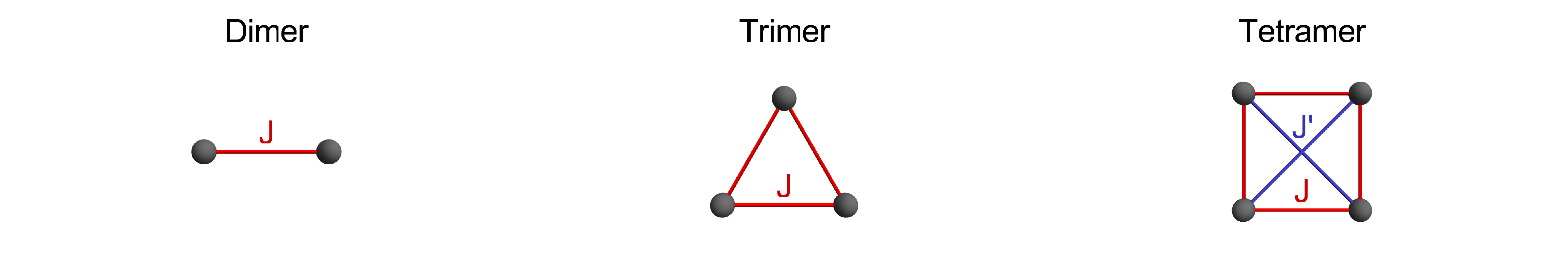}
		\includegraphics[width=1\textwidth]{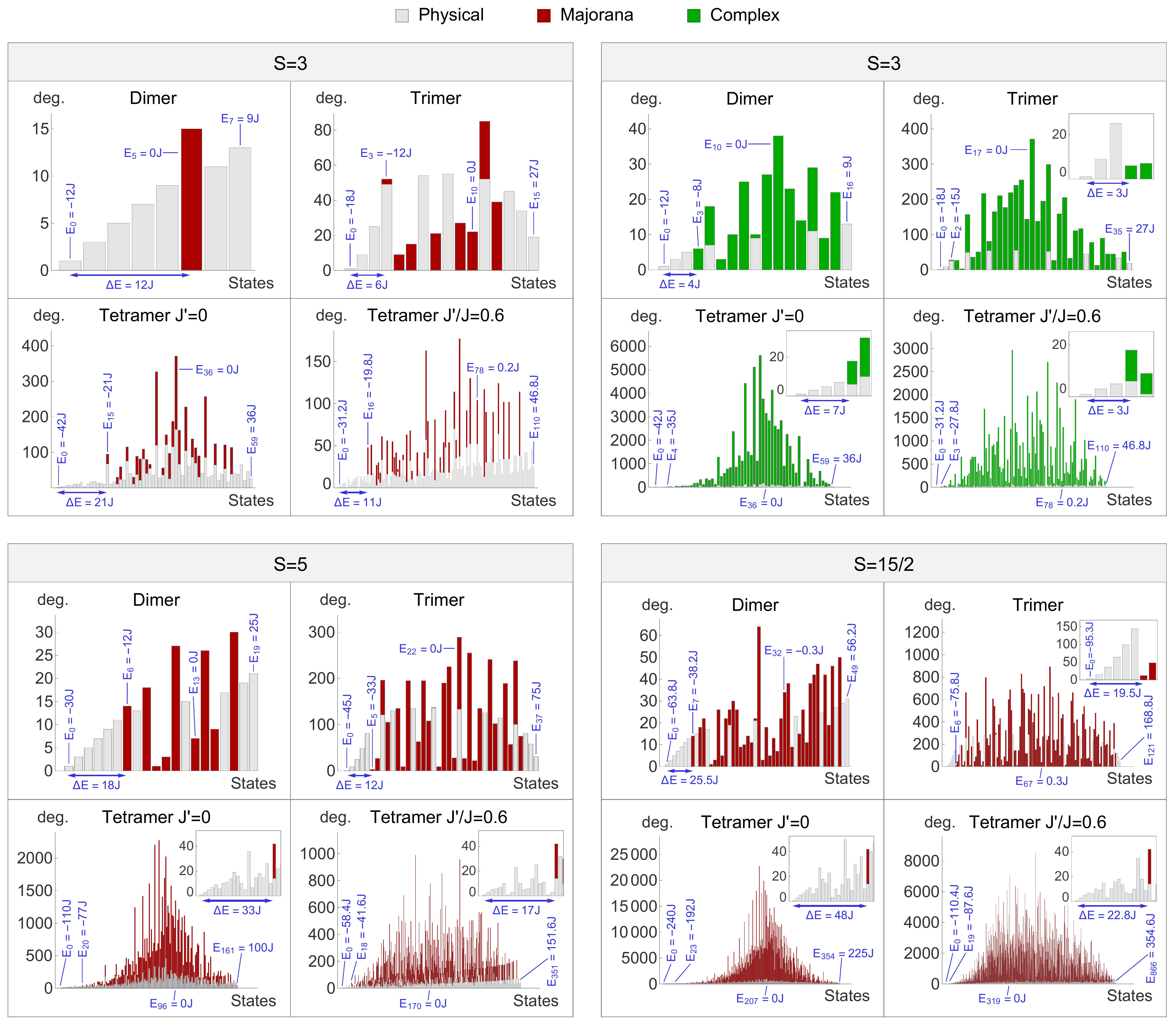}
	\caption[States of spin clusters]{Histograms for the numbers of degenerate states for the spin dimer, trimer, and tetramer, which are illustrated in the top panel (where the tetramer is considered for $J'=0$ and $J'/J=0.6$). In the four main panels the spin amplitudes are given by $S=3,5,\frac{15}{2}$. Gray states correspond to the physical states obtained with exact spin operators. The red states are additional unphysical states from extra spin sectors, which result from our Majorana spin representations with $s=3,4,5$, respectively. The green states are additional unphysical states with $S<3$ occurring for a complex fermionic representation with $s=2$. Note that the histograms show the spin states ordered by their energy, however, the horizontral axes are not linear in energy. The energy difference $\Delta E$ between the ground states in the physical and unphysical sectors are indicated in each plot or in the insets showing magnified versions of the low energy regimes. The energies of some selected states are explicitly given.}
	\label{fig:ClustersStates}
	\end{center}
\end{figure*}

\begin{figure*}
	\begin{center}
		\includegraphics[width=1.0\textwidth]{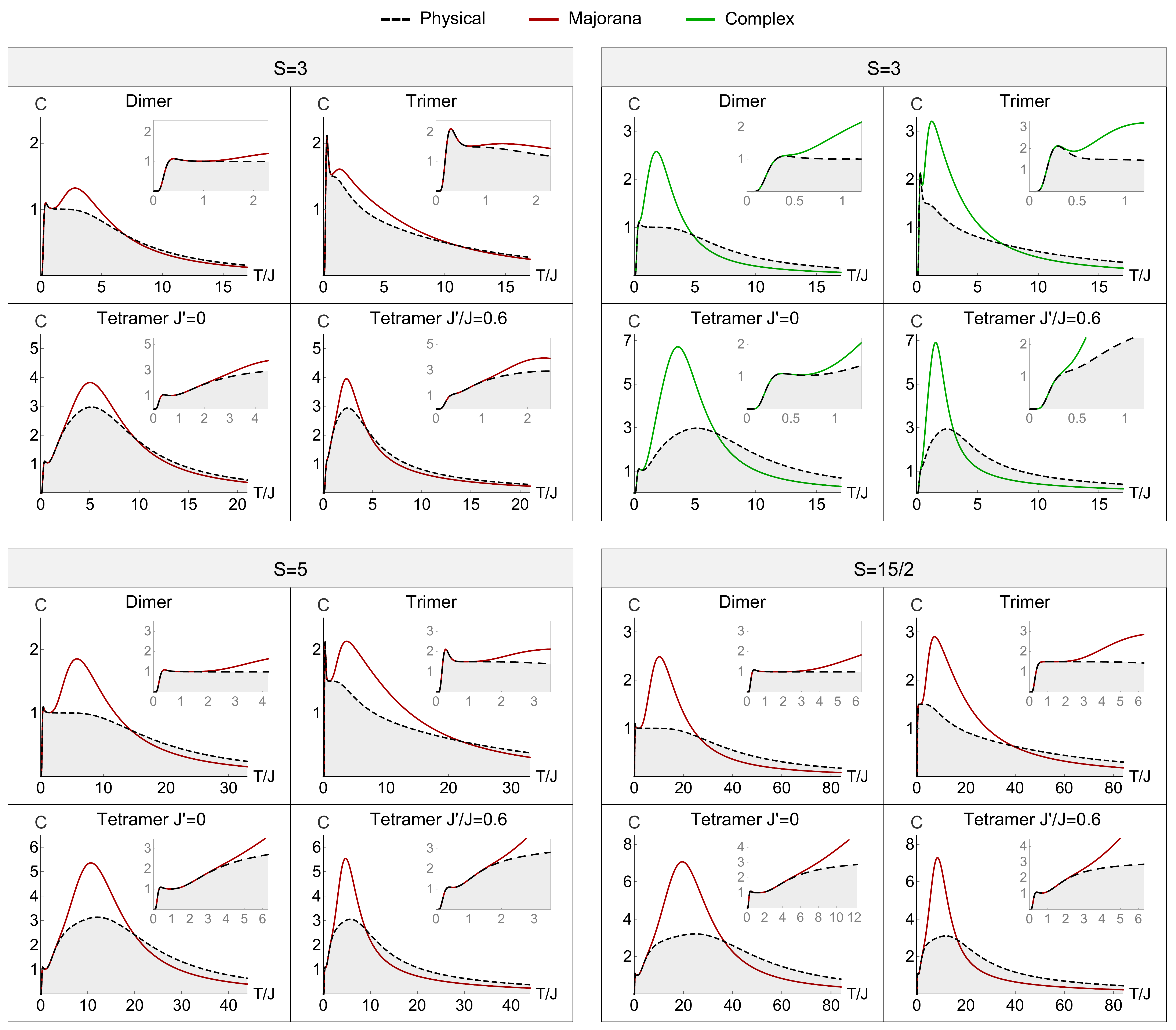}
	\caption[Heat Capacities of Spin Clusters]{The heat capacity $C(T)$ as a function of temperature $T$ for the spin clusters dimer, trimer and tetramer for spins $S=3$, 5 and $\frac{15}{2}$. The temperature is given in units of the nearest neighbor coupling constant $J$. For the tetramer we consider the two cases, where the interaction $J'$ between second nearest neighbors is either zero or $J'/J=0.6$. Black, dashed graphs illustrate the heat capacity computed from physical states only. Graphs shown in red (green) are calculated by taking into account the additional states from the Majorana (complex fermionic) representation. The insets show the heat capacity at low temperatures.}
	\label{fig:HeatCapacities}
	\end{center}
\end{figure*}

\subsection{Majorana spin representations with $s>3$} \label{sec:higherspin}
While for Majorana spin representations with $s>3$, the number of additional subspaces with $S<S_\text{max}$ quickly increases, occupying large parts of the Hilbert space, it is interesting to observe that the second largest realized spin magnitude is always $S=S_\text{max}-3$ (as has already been found in the previous Sec.~\ref{sec:spin3} for the special case $s=3$). This can be seen from the set of generated magnetic quantum numbers $M^z$ in Eq.~(\ref{eq:mzmajo}) where -- apart from the double degeneracy from $\lambda_0=\pm1$ -- there is a unique combinatorial choice of eigenvalues $\lambda_m=\pm1$ with $m\in\{1,\ldots,s\}$ which yields the values $M^z=S_\text{max}$, $M^z=S_\text{max}-1$, and $M^z=S_\text{max}-2$ (note that, to obtain $M^z=S_\text{max}$ all eigenvalues must be $\lambda_m=+1$). However, the value $M^z=S_\text{max}-3$ can be obtained with two different sign combinations, namely
\begin{align}
    S_\text{max}-3&=-\frac{1}{2}-1+\frac{3}{2}+2+\ldots\\
    &=+\frac{1}{2}+1-\frac{3}{2}+2+\ldots,
\end{align}
which means that another spin space with $S=S_\text{max}-3$ must exist. This difference of three units of angular momentum between the largest and second largest spin magnitude can be considered advantageous compared to complex fermionic representations where, e.g., for $s=3/2$ ($s=2$) this difference is only one half (one) unit of angular momentum, see Appendix. Continuing the above argument for further spin subspaces realized by our Majorana representations, one finds that for $s>4$ the spin space with the third largest spin amplitude has $S=S_\text{max}-5$.  

\section{Influence of additional spin sectors} \label{sec:states}
In this section, we apply our Majorana spin representations from Eqs.~(\ref{majorepnx})-(\ref{majorepnz}) to small clusters of interacting spins that can be diagonalized exactly. We, intentionally, do not apply any projection scheme on the states to be able to asses the impact of different spin sectors on the energy spectrum. We do neglect the identical copy of each spin spectrum (see Table~\ref{tab1}) though, as it always appears in the Majorana representation and leads to an increasing redundancy in larger spin clusters. More specifically, we consider Majorana representations for $s=3,4,5$ which realize the spin amplitudes $S=3,5,\frac{15}{2}$ in the largest spin sector, respectively, [see Eq.~(\ref{eq:smax})]. In the following, we define this sector as the ``physical'' one. Our clusters of equal spins consist of two spins (dimer), three spins (trimer) and four spins (tetramer) which interact via antiferromagnetic ($J_{ij}>0$) Heisenberg couplings
\begin{equation}
    H=\sum_{i>j}J_{ij}\bm{S}_i\bm{S}_j.
\end{equation}
As illustrated in the top part of Fig.~\ref{fig:ClustersStates}, the interactions $J_{ij}$ are all equal for the trimer ($J_{ij}\equiv J$), while for the tetramer we consider two cases, where the diagonal couplings $J'$ either vanish ($J'=0$) or where the diagonal and nearest neighbor couplings $J$ are in the ratio of $J'/J=0.6$. These clusters represent typical building blocks of frustrated spin systems such as kagome, $J_1$-$J_2$ square, and pyrochlore lattices.

For all considered clusters and spin amplitudes the ground states and a few low energy states are found to reside in the physical sector (see gray states in Fig.~\ref{fig:ClustersStates}). The lowest unphysical state (i.e., lowest red state) is separated from the ground state by an energy difference $\Delta E$ which, on average over all clusters and spin amplitudes, is approximately given by $\approx6J$ {\it per site}, however, this number varies considerably between the different systems. We also find a robust energy difference between the highest physical and unphysical states at the upper end of the spectrum, indicating that the same conclusions also hold in the case of ferromagnetic couplings. Furthermore, for the tetramer with the largest considered spin $S=\frac{15}{2}$ the histograms in Fig.~\ref{fig:ClustersStates} illustrate how quickly the spectrum is populated with unphysical states. This is because a single spin in an unphysical sector is already sufficient to characterize the full spin state as unphysical.

It is interesting to compare these findings with corresponding results from the complex fermionic representations in Eq.~(\ref{eq:complex_rep1}). For example, in contrast to the trimer spectra in Fig.~\ref{fig:ClustersStates}, the spin-$1/2$ trimer in Abrikosov representation has degenerate ground states in the physical {\it and} unphysical sectors, as has recently been discussed in Ref.~\cite{schneider22}. Since the Majorana representation with $s=3$ and the complex fermionic representation with $s=2$ both realize a spin $S=3$ in the largest spin sector, this case is particularly suited for a direct comparison between the two representations, see the red and green histograms in the two panels for $S=3$ in Fig.~\ref{fig:ClustersStates}. The spectra for the complex fermionic representation in green exhibit a larger number of unphysical states and show a significantly reduced energy difference $\Delta E$ between the lowest states in the physical and unphysical sectors. Both observations are a direct consequence of the more complicated structure of spin spaces in the complex fermionic representation and the smaller difference in spin amplitudes between physical and unphysical sectors which is only $\Delta S=1$ in this case. These results demonstrate that the rather large value $\Delta S=3$ for the Majorana representations yields a clear advantage in separating physical and unphysical states at low energies. Whether these properties also remain for larger spin systems, however, cannot be extrapolated based on the current results.

Finally, we discuss how the unphysical states in these spin clusters affect thermodynamic quantities, such as the heat capacity $C(T)=\partial \langle E\rangle/\partial T$ which we show in Fig.~\ref{fig:HeatCapacities} as a function of temperature $T$ for the same clusters (dimer, trimer and tetramer) and spin amplitudes ($S=3,5,\frac{15}{2}$) as in Fig.~\ref{fig:ClustersStates}. Comparing the heat capacities computed for exact spin operators (i.e., without any unphysical states) and with the Majorana representations shows that in all considered cases, the unphysical states have a negligible impact on $C(T)$ at least up to temperatures $T\approx J$. For the tetramer and the largest considered spin amplitudes the temperature region where the Majorana representations yield accurate results ranges up to even larger temperatures $T\approx2J\ldots 6J$. On the other hand, for the trimer, the heat capacities from exact spins and Majorana representations already deviate at comparatively smaller temperatures. Furthermore, in accordance with the observations in Fig.~\ref{fig:ClustersStates}, the heat capacities for spin-$3$ clusters in complex fermionic representation start differing from the exact results already significantly below $T=J$.

\section{Summary and conclusion}\label{sec:summary}
In summary, we have shown that bilinear Majorana representations of spin operators can be constructed systematically from the real irreducible matrix representations of SU(2). We presented two methods to derive these matrix representations from the higher spin ($s>1/2$) Pauli matrices. One approach was based on their realification which resulted in Majorana representations that are equivalent to the known complex fermionic spin representations~\cite{liu10}. The other method made use of the direct unitary transformation of odd-dimensional Pauli matrices which uncovered a family of bilinear Majorana representations for spin operators with $S=s(s+1)/4$ where $s\in\mathbb{N}$. We derived closed analytical expressions for the latter ones, which reproduce the known SO(3) Majorana representation for spin-$1/2$~\cite{martin59,tsvelik92} as well as a spin-$3/2$ Majorana representation~\cite{yao09}. While these two representations do not introduce any additional spin spaces (apart from a two-fold redundancy) this is no longer the case for larger spin amplitudes $S>3/2$. We explained that these additional spin spaces cannot be easily projected out because no bilinear Majorana operator can distinguish them. As a consequence, the application of an extension of the Popov-Fedotov projection scheme is significantly more complicated. On the other hand, however, the intermixing of different spin spaces is less severe compared to complex fermionic representations. For example, the largest and second largest spin amplitudes are well separated by a difference of $\Delta S=3$. These advantages are particularly obvious for our spin-$3$ Majorana representation where, for a single spin, the dimensions of physical and unphysical spin spaces are in a ratio of $7:1$. Even though this ratio quickly decreases as more spins are added, our preliminary results for small spin clusters nevertheless indicate that the low-energy properties are not effected by unphysical states. Whether this property holds true for larger clusters or even in the thermodynamic limit of a spin model can, however, not be decided based on the current results. A spin-$3$ degree of freedom may already be considered as well suited to approximate the classical large spin limit. Hence, applying these representations within the PMFRG-type approaches mentioned at the beginning of this work, will open up interesting perspectives to numerically explore the fate of a quantum spin system upon approaching the limit of classical spins.

\section{Acknowledgement}
We thank Nils Niggemann, Bj\"orn Sbierski, and Benedikt Schneider for insightful discussions and acknowledge support from the Deutsche Forschungsgemeinschaft (DFG, German Research Foundation) - Project Number 277101999 - CRC 183 (project A04). J.\,R. thanks IIT Madras for a Visiting Faculty Fellow position under the IoE program during which part of the research work was carried out.

\appendix

\subsection*{Appendix: Generated spin spaces for Majorana and complex fermionic spin representations}
In Table~\ref{tab5}, we list the spin spaces generated in our Majorana representations [see Eqs.~(\ref{majorepnx})-(\ref{majorepnz})] and in the complex fermionic representations [see Eq.~(\ref{eq:complex_rep1})] for all values $s\leq5$.

\begin{table*}[t]
	\footnotesize
	\renewcommand\arraystretch{1.3}
	\setlength\tabcolsep{0.15cm}
	\begin{tabular}{|c|c|l|}
		\hline
		$\boldsymbol{s}$ & \textbf{Repn} & \textbf{Generated spin spaces} \\
		\hline
		\hline\rule{0pt}{11pt}
		1/2& Complex & $\frac{1}{2} \oplus 0^2$ \\[2pt]
		\hline
		\multirow{ 2}{*}{1} & Majorana & $\frac{1}{2} \oplus \mathrm{copies}$  \\
		& Complex & $1 \oplus 0 \oplus \mathrm{copies}$ \\
		\hline\rule{0pt}{11pt}
		3/2& Complex & $0^3 \oplus \frac{3}{2}^2 \oplus 2$ \\[2pt]
		\hline
		\multirow{ 2}{*}{2}& Majorana & $\frac{3}{2} \oplus \mathrm{copies}$  \\
		& Complex & $3 \oplus 2 \oplus 1 \oplus 0 \oplus \mathrm{copies}$ \\
			\hline\rule{0pt}{11pt}
		5/2& Complex & $\frac{9}{2} \oplus 4^2 \oplus \frac{5}{2}^3 \oplus 2^2 \oplus \frac{3}{2} \oplus 0^4$ \\[2pt]
		\hline
		\multirow{ 2}{*}{3}& Majorana & $3 \oplus 0 \oplus \mathrm{copies}$ \\
		& Complex & $6 \oplus 5 \oplus 4 \oplus 3^3 \oplus 2 \oplus 1 \oplus 0^2 \oplus \mathrm{copies}$ \\
		\hline\rule{0pt}{11pt}
		7/2& Complex & $8 \oplus \frac{15}{2}^2 \oplus 6^3 \oplus \frac{11}{2}^2 \oplus 5 \oplus \frac{9}{2}^2 \oplus 4^4 \oplus \frac{7}{2}^4 \oplus \frac{5}{2}^2 \oplus 2^4 \oplus \frac{3}{2}^2 \oplus 0^5$ \\[2pt]
		\hline
		\multirow{ 2}{*}{4}& Majorana & $5 \oplus 2 \oplus \mathrm{copies}$ \\
		& Complex & $10 \oplus 9 \oplus 8 \oplus 7^3 \oplus 6^2 \oplus 5^3 \oplus 4^4 \oplus 3^4 \oplus 2^2 \oplus 1^2 \oplus 0^2 \oplus \mathrm{copies}$ \\
		\hline\rule{0pt}{11pt}
		9/2& Complex & $\frac{25}{2}\1\oplus\1 12^2\1 \oplus\1 \frac{21}{2}^3 \1\oplus\1 10^2 \1\oplus\1 \frac{19}{2} \1\oplus\1 9^2 \1\oplus\1 \frac{17}{2}^4 \1\oplus\1 8^6 \1\oplus\1 \frac{15}{2}^4 \1\oplus\1 7^2 \1\oplus\1 \frac{13}{2}^4 \1\oplus\1 6^8 \1\oplus\1 \frac{11}{2}^4 \1\oplus\1 5^2 \1\oplus\1 \frac{9}{2}^9 \1\oplus\1 4^8 \1\oplus\1 \frac{7}{2}^4 \1\oplus\1 3^2 \1\oplus\1 \frac{5}{2}^4 \1\oplus\1 2^6 \1\oplus\1 \frac{3}{2}^3 \1\oplus\1 \frac{1}{2} \1\oplus\1 0^8$ \\[2pt]
		\hline
		\multirow{ 2}{*}{5}& Majorana & $\frac{15}{2} \oplus \frac{9}{2} \oplus \frac{5}{2} \oplus \mathrm{copies}$ \\
		& Complex & $15 \oplus 14 \oplus 13 \oplus 12^3 \oplus 11^3 \oplus 10^5 \oplus 9^6 \oplus 8^6 \oplus 7^8 \oplus 6^8 \oplus 5^9 \oplus 4^7 \oplus 3^7 \oplus 2^5 \oplus 1^3 \oplus 0^3 \oplus \mathrm{copies}$ \\
		\hline
	\end{tabular}
	\caption{Spin spaces, generated by the Majorana representations in Eqs.~(\ref{majorepnx})-(\ref{majorepnz}) and the complex fermionic representations in Eq.~(\ref{eq:complex_rep1}) for different values of $s$. Exponents denote the degeneracy of the respective spin space. Note that the Majorana representations considered here only exist for integer $s$, i.e. odd $d=2s+1$.}
	\label{tab5}
\end{table*}


\begin{thebibliography}{75}%
\makeatletter
\providecommand \@ifxundefined [1]{%
 \@ifx{#1\undefined}
}%
\providecommand \@ifnum [1]{%
 \ifnum #1\expandafter \@firstoftwo
 \else \expandafter \@secondoftwo
 \fi
}%
\providecommand \@ifx [1]{%
 \ifx #1\expandafter \@firstoftwo
 \else \expandafter \@secondoftwo
 \fi
}%
\providecommand \natexlab [1]{#1}%
\providecommand \enquote  [1]{``#1''}%
\providecommand \bibnamefont  [1]{#1}%
\providecommand \bibfnamefont [1]{#1}%
\providecommand \citenamefont [1]{#1}%
\providecommand \href@noop [0]{\@secondoftwo}%
\providecommand \href [0]{\begingroup \@sanitize@url \@href}%
\providecommand \@href[1]{\@@startlink{#1}\@@href}%
\providecommand \@@href[1]{\endgroup#1\@@endlink}%
\providecommand \@sanitize@url [0]{\catcode `\\12\catcode `\$12\catcode
  `\&12\catcode `\#12\catcode `\^12\catcode `\_12\catcode `\%12\relax}%
\providecommand \@@startlink[1]{}%
\providecommand \@@endlink[0]{}%
\providecommand \url  [0]{\begingroup\@sanitize@url \@url }%
\providecommand \@url [1]{\endgroup\@href {#1}{\urlprefix }}%
\providecommand \urlprefix  [0]{URL }%
\providecommand \Eprint [0]{\href }%
\providecommand \doibase [0]{http://dx.doi.org/}%
\providecommand \selectlanguage [0]{\@gobble}%
\providecommand \bibinfo  [0]{\@secondoftwo}%
\providecommand \bibfield  [0]{\@secondoftwo}%
\providecommand \translation [1]{[#1]}%
\providecommand \BibitemOpen [0]{}%
\providecommand \bibitemStop [0]{}%
\providecommand \bibitemNoStop [0]{.\EOS\space}%
\providecommand \EOS [0]{\spacefactor3000\relax}%
\providecommand \BibitemShut  [1]{\csname bibitem#1\endcsname}%
\let\auto@bib@innerbib\@empty
\bibitem [{\citenamefont {Jordan}\ and\ \citenamefont
  {Wigner}(1928)}]{jordan28}%
  \BibitemOpen
  \bibfield  {author} {\bibinfo {author} {\bibfnamefont {P.}~\bibnamefont
  {Jordan}}\ and\ \bibinfo {author} {\bibfnamefont {E.}~\bibnamefont
  {Wigner}},\ }\href {\doibase 10.1007/BF01331938} {\bibfield  {journal}
  {\bibinfo  {journal} {Zeitschrift f{\"u}r Physik}\ }\textbf {\bibinfo
  {volume} {47}},\ \bibinfo {pages} {631} (\bibinfo {year} {1928})}\BibitemShut
  {NoStop}%
\bibitem [{\citenamefont {Holstein}\ and\ \citenamefont
  {Primakoff}(1940)}]{holstein40}%
  \BibitemOpen
  \bibfield  {author} {\bibinfo {author} {\bibfnamefont {T.}~\bibnamefont
  {Holstein}}\ and\ \bibinfo {author} {\bibfnamefont {H.}~\bibnamefont
  {Primakoff}},\ }\href {\doibase 10.1103/PhysRev.58.1098} {\bibfield
  {journal} {\bibinfo  {journal} {Phys. Rev.}\ }\textbf {\bibinfo {volume}
  {58}},\ \bibinfo {pages} {1098} (\bibinfo {year} {1940})}\BibitemShut
  {NoStop}%
\bibitem [{\citenamefont {Anderson}(1952)}]{anderson52}%
  \BibitemOpen
  \bibfield  {author} {\bibinfo {author} {\bibfnamefont {P.~W.}\ \bibnamefont
  {Anderson}},\ }\href {\doibase 10.1103/PhysRev.86.694} {\bibfield  {journal}
  {\bibinfo  {journal} {Phys. Rev.}\ }\textbf {\bibinfo {volume} {86}},\
  \bibinfo {pages} {694} (\bibinfo {year} {1952})}\BibitemShut {NoStop}%
\bibitem [{\citenamefont {Kubo}(1952)}]{kubo52}%
  \BibitemOpen
  \bibfield  {author} {\bibinfo {author} {\bibfnamefont {R.}~\bibnamefont
  {Kubo}},\ }\href {\doibase 10.1103/PhysRev.87.568} {\bibfield  {journal}
  {\bibinfo  {journal} {Phys. Rev.}\ }\textbf {\bibinfo {volume} {87}},\
  \bibinfo {pages} {568} (\bibinfo {year} {1952})}\BibitemShut {NoStop}%
\bibitem [{\citenamefont {Abrikosov}(1965)}]{abrikosov_electron_1965}%
  \BibitemOpen
  \bibfield  {author} {\bibinfo {author} {\bibfnamefont {A.~A.}\ \bibnamefont
  {Abrikosov}},\ }\href {\doibase 10.1103/PhysicsPhysiqueFizika.2.5} {\bibfield
   {journal} {\bibinfo  {journal} {Physics Physique Fizika}\ }\textbf {\bibinfo
  {volume} {2}},\ \bibinfo {pages} {5} (\bibinfo {year} {1965})}\BibitemShut
  {NoStop}%
\bibitem [{\citenamefont {Zhang}\ and\ \citenamefont {Wang}(2006)}]{zhang06}%
  \BibitemOpen
  \bibfield  {author} {\bibinfo {author} {\bibfnamefont {G.-M.}\ \bibnamefont
  {Zhang}}\ and\ \bibinfo {author} {\bibfnamefont {X.}~\bibnamefont {Wang}},\
  }\href {\doibase 10.1088/0305-4470/39/26/017} {\bibfield  {journal} {\bibinfo
   {journal} {Journal of Physics A: Mathematical and General}\ }\textbf
  {\bibinfo {volume} {39}},\ \bibinfo {pages} {8515} (\bibinfo {year}
  {2006})}\BibitemShut {NoStop}%
\bibitem [{\citenamefont {Tu}\ \emph {et~al.}(2009)\citenamefont {Tu},
  \citenamefont {Zhang}, \citenamefont {Xiang}, \citenamefont {Liu},\ and\
  \citenamefont {Ng}}]{hong09}%
  \BibitemOpen
  \bibfield  {author} {\bibinfo {author} {\bibfnamefont {H.-H.}\ \bibnamefont
  {Tu}}, \bibinfo {author} {\bibfnamefont {G.-M.}\ \bibnamefont {Zhang}},
  \bibinfo {author} {\bibfnamefont {T.}~\bibnamefont {Xiang}}, \bibinfo
  {author} {\bibfnamefont {Z.-X.}\ \bibnamefont {Liu}}, \ and\ \bibinfo
  {author} {\bibfnamefont {T.-K.}\ \bibnamefont {Ng}},\ }\href {\doibase
  10.1103/PhysRevB.80.014401} {\bibfield  {journal} {\bibinfo  {journal} {Phys.
  Rev. B}\ }\textbf {\bibinfo {volume} {80}},\ \bibinfo {pages} {014401}
  (\bibinfo {year} {2009})}\BibitemShut {NoStop}%
\bibitem [{\citenamefont {Liu}\ \emph {et~al.}(2010)\citenamefont {Liu},
  \citenamefont {Zhou},\ and\ \citenamefont {Ng}}]{liu10}%
  \BibitemOpen
  \bibfield  {author} {\bibinfo {author} {\bibfnamefont {Z.-X.}\ \bibnamefont
  {Liu}}, \bibinfo {author} {\bibfnamefont {Y.}~\bibnamefont {Zhou}}, \ and\
  \bibinfo {author} {\bibfnamefont {T.-K.}\ \bibnamefont {Ng}},\ }\href
  {\doibase 10.1103/PhysRevB.82.144422} {\bibfield  {journal} {\bibinfo
  {journal} {Phys. Rev. B}\ }\textbf {\bibinfo {volume} {82}},\ \bibinfo
  {pages} {144422} (\bibinfo {year} {2010})}\BibitemShut {NoStop}%
\bibitem [{\citenamefont {Savary}\ and\ \citenamefont
  {Balents}(2017)}]{savary_quantum_2017}%
  \BibitemOpen
  \bibfield  {author} {\bibinfo {author} {\bibfnamefont {L.}~\bibnamefont
  {Savary}}\ and\ \bibinfo {author} {\bibfnamefont {L.}~\bibnamefont
  {Balents}},\ }\href {\doibase 10.1088/0034-4885/80/1/016502} {\bibfield
  {journal} {\bibinfo  {journal} {Rep. Prog. Phys.}\ }\textbf {\bibinfo
  {volume} {80}},\ \bibinfo {pages} {016502} (\bibinfo {year}
  {2017})}\BibitemShut {NoStop}%
\bibitem [{\citenamefont {Affleck}\ and\ \citenamefont
  {Marston}(1988)}]{affleck88}%
  \BibitemOpen
  \bibfield  {author} {\bibinfo {author} {\bibfnamefont {I.}~\bibnamefont
  {Affleck}}\ and\ \bibinfo {author} {\bibfnamefont {J.~B.}\ \bibnamefont
  {Marston}},\ }\href {\doibase 10.1103/PhysRevB.37.3774} {\bibfield  {journal}
  {\bibinfo  {journal} {Phys. Rev. B}\ }\textbf {\bibinfo {volume} {37}},\
  \bibinfo {pages} {3774} (\bibinfo {year} {1988})}\BibitemShut {NoStop}%
\bibitem [{\citenamefont {Wen}(1991)}]{wen91}%
  \BibitemOpen
  \bibfield  {author} {\bibinfo {author} {\bibfnamefont {X.~G.}\ \bibnamefont
  {Wen}},\ }\href {\doibase 10.1103/PhysRevB.44.2664} {\bibfield  {journal}
  {\bibinfo  {journal} {Phys. Rev. B}\ }\textbf {\bibinfo {volume} {44}},\
  \bibinfo {pages} {2664} (\bibinfo {year} {1991})}\BibitemShut {NoStop}%
\bibitem [{\citenamefont {Wen}(2002)}]{wen02}%
  \BibitemOpen
  \bibfield  {author} {\bibinfo {author} {\bibfnamefont {X.-G.}\ \bibnamefont
  {Wen}},\ }\href {\doibase 10.1103/PhysRevB.65.165113} {\bibfield  {journal}
  {\bibinfo  {journal} {Phys. Rev. B}\ }\textbf {\bibinfo {volume} {65}},\
  \bibinfo {pages} {165113} (\bibinfo {year} {2002})}\BibitemShut {NoStop}%
\bibitem [{\citenamefont {Wen}(2004)}]{wen_book}%
  \BibitemOpen
  \bibfield  {author} {\bibinfo {author} {\bibfnamefont {X.-G.}\ \bibnamefont
  {Wen}},\ }\href@noop {} {\emph {\bibinfo {title} {Quantum field theory of
  many-body systems: From the origin of sound to an origin of light and
  electrons}}}\ (\bibinfo  {publisher} {Oxford University Press},\ \bibinfo
  {year} {2004})\BibitemShut {NoStop}%
\bibitem [{\citenamefont {Reuther}\ and\ \citenamefont
  {W\"olfle}(2010)}]{reuther_j_2010}%
  \BibitemOpen
  \bibfield  {author} {\bibinfo {author} {\bibfnamefont {J.}~\bibnamefont
  {Reuther}}\ and\ \bibinfo {author} {\bibfnamefont {P.}~\bibnamefont
  {W\"olfle}},\ }\href {\doibase 10.1103/PhysRevB.81.144410} {\bibfield
  {journal} {\bibinfo  {journal} {Phys. Rev. B}\ }\textbf {\bibinfo {volume}
  {81}},\ \bibinfo {pages} {144410} (\bibinfo {year} {2010})}\BibitemShut
  {NoStop}%
\bibitem [{\citenamefont {Iqbal}\ \emph {et~al.}(2016)\citenamefont {Iqbal},
  \citenamefont {Thomale}, \citenamefont {Parisen~Toldin}, \citenamefont
  {Rachel},\ and\ \citenamefont {Reuther}}]{iqbal16}%
  \BibitemOpen
  \bibfield  {author} {\bibinfo {author} {\bibfnamefont {Y.}~\bibnamefont
  {Iqbal}}, \bibinfo {author} {\bibfnamefont {R.}~\bibnamefont {Thomale}},
  \bibinfo {author} {\bibfnamefont {F.}~\bibnamefont {Parisen~Toldin}},
  \bibinfo {author} {\bibfnamefont {S.}~\bibnamefont {Rachel}}, \ and\ \bibinfo
  {author} {\bibfnamefont {J.}~\bibnamefont {Reuther}},\ }\href {\doibase
  10.1103/PhysRevB.94.140408} {\bibfield  {journal} {\bibinfo  {journal} {Phys.
  Rev. B}\ }\textbf {\bibinfo {volume} {94}},\ \bibinfo {pages} {140408}
  (\bibinfo {year} {2016})}\BibitemShut {NoStop}%
\bibitem [{\citenamefont {Baez}\ and\ \citenamefont
  {Reuther}(2017)}]{baez_numerical_2017}%
  \BibitemOpen
  \bibfield  {author} {\bibinfo {author} {\bibfnamefont {M.~L.}\ \bibnamefont
  {Baez}}\ and\ \bibinfo {author} {\bibfnamefont {J.}~\bibnamefont {Reuther}},\
  }\href {\doibase 10.1103/PhysRevB.96.045144} {\bibfield  {journal} {\bibinfo
  {journal} {Phys. Rev. B}\ }\textbf {\bibinfo {volume} {96}},\ \bibinfo
  {pages} {045144} (\bibinfo {year} {2017})}\BibitemShut {NoStop}%
\bibitem [{\citenamefont {R\"uck}\ and\ \citenamefont
  {Reuther}(2018)}]{rueck18}%
  \BibitemOpen
  \bibfield  {author} {\bibinfo {author} {\bibfnamefont {M.}~\bibnamefont
  {R\"uck}}\ and\ \bibinfo {author} {\bibfnamefont {J.}~\bibnamefont
  {Reuther}},\ }\href {\doibase 10.1103/PhysRevB.97.144404} {\bibfield
  {journal} {\bibinfo  {journal} {Phys. Rev. B}\ }\textbf {\bibinfo {volume}
  {97}},\ \bibinfo {pages} {144404} (\bibinfo {year} {2018})}\BibitemShut
  {NoStop}%
\bibitem [{\citenamefont {Roscher}\ \emph {et~al.}(2018)\citenamefont
  {Roscher}, \citenamefont {Buessen}, \citenamefont {Scherer}, \citenamefont
  {Trebst},\ and\ \citenamefont {Diehl}}]{roscher18}%
  \BibitemOpen
  \bibfield  {author} {\bibinfo {author} {\bibfnamefont {D.}~\bibnamefont
  {Roscher}}, \bibinfo {author} {\bibfnamefont {F.~L.}\ \bibnamefont
  {Buessen}}, \bibinfo {author} {\bibfnamefont {M.~M.}\ \bibnamefont
  {Scherer}}, \bibinfo {author} {\bibfnamefont {S.}~\bibnamefont {Trebst}}, \
  and\ \bibinfo {author} {\bibfnamefont {S.}~\bibnamefont {Diehl}},\ }\href
  {\doibase 10.1103/PhysRevB.97.064416} {\bibfield  {journal} {\bibinfo
  {journal} {Phys. Rev. B}\ }\textbf {\bibinfo {volume} {97}},\ \bibinfo
  {pages} {064416} (\bibinfo {year} {2018})}\BibitemShut {NoStop}%
\bibitem [{\citenamefont {Buessen}\ \emph {et~al.}(2018)\citenamefont
  {Buessen}, \citenamefont {Roscher}, \citenamefont {Diehl},\ and\
  \citenamefont {Trebst}}]{buessen18}%
  \BibitemOpen
  \bibfield  {author} {\bibinfo {author} {\bibfnamefont {F.~L.}\ \bibnamefont
  {Buessen}}, \bibinfo {author} {\bibfnamefont {D.}~\bibnamefont {Roscher}},
  \bibinfo {author} {\bibfnamefont {S.}~\bibnamefont {Diehl}}, \ and\ \bibinfo
  {author} {\bibfnamefont {S.}~\bibnamefont {Trebst}},\ }\href {\doibase
  10.1103/PhysRevB.97.064415} {\bibfield  {journal} {\bibinfo  {journal} {Phys.
  Rev. B}\ }\textbf {\bibinfo {volume} {97}},\ \bibinfo {pages} {064415}
  (\bibinfo {year} {2018})}\BibitemShut {NoStop}%
\bibitem [{\citenamefont {Buessen}\ \emph {et~al.}(2019)\citenamefont
  {Buessen}, \citenamefont {Noculak}, \citenamefont {Trebst},\ and\
  \citenamefont {Reuther}}]{buessen19}%
  \BibitemOpen
  \bibfield  {author} {\bibinfo {author} {\bibfnamefont {F.~L.}\ \bibnamefont
  {Buessen}}, \bibinfo {author} {\bibfnamefont {V.}~\bibnamefont {Noculak}},
  \bibinfo {author} {\bibfnamefont {S.}~\bibnamefont {Trebst}}, \ and\ \bibinfo
  {author} {\bibfnamefont {J.}~\bibnamefont {Reuther}},\ }\href {\doibase
  10.1103/PhysRevB.100.125164} {\bibfield  {journal} {\bibinfo  {journal}
  {Phys. Rev. B}\ }\textbf {\bibinfo {volume} {100}},\ \bibinfo {pages}
  {125164} (\bibinfo {year} {2019})}\BibitemShut {NoStop}%
\bibitem [{\citenamefont {Buessen}(2022)}]{buessen_code}%
  \BibitemOpen
  \bibfield  {author} {\bibinfo {author} {\bibfnamefont {F.~L.}\ \bibnamefont
  {Buessen}},\ }\href {\doibase 10.21468/SciPostPhysCodeb.5} {\bibfield
  {journal} {\bibinfo  {journal} {SciPost Phys. Codebases}\ ,\ \bibinfo {pages}
  {5}} (\bibinfo {year} {2022})}\BibitemShut {NoStop}%
\bibitem [{\citenamefont {Kiese}\ \emph {et~al.}(2020)\citenamefont {Kiese},
  \citenamefont {Buessen}, \citenamefont {Hickey}, \citenamefont {Trebst},\
  and\ \citenamefont {Scherer}}]{kiese20}%
  \BibitemOpen
  \bibfield  {author} {\bibinfo {author} {\bibfnamefont {D.}~\bibnamefont
  {Kiese}}, \bibinfo {author} {\bibfnamefont {F.~L.}\ \bibnamefont {Buessen}},
  \bibinfo {author} {\bibfnamefont {C.}~\bibnamefont {Hickey}}, \bibinfo
  {author} {\bibfnamefont {S.}~\bibnamefont {Trebst}}, \ and\ \bibinfo {author}
  {\bibfnamefont {M.~M.}\ \bibnamefont {Scherer}},\ }\href {\doibase
  10.1103/PhysRevResearch.2.013370} {\bibfield  {journal} {\bibinfo  {journal}
  {Phys. Rev. Research}\ }\textbf {\bibinfo {volume} {2}},\ \bibinfo {pages}
  {013370} (\bibinfo {year} {2020})}\BibitemShut {NoStop}%
\bibitem [{\citenamefont {Thoenniss}\ \emph {et~al.}(2020)\citenamefont
  {Thoenniss}, \citenamefont {Ritter}, \citenamefont {Kugler}, \citenamefont
  {{von Delft}},\ and\ \citenamefont {Punk}}]{thoenniss20}%
  \BibitemOpen
  \bibfield  {author} {\bibinfo {author} {\bibfnamefont {J.}~\bibnamefont
  {Thoenniss}}, \bibinfo {author} {\bibfnamefont {M.~K.}\ \bibnamefont
  {Ritter}}, \bibinfo {author} {\bibfnamefont {F.~B.}\ \bibnamefont {Kugler}},
  \bibinfo {author} {\bibfnamefont {J.}~\bibnamefont {{von Delft}}}, \ and\
  \bibinfo {author} {\bibfnamefont {M.}~\bibnamefont {Punk}},\ }\href@noop {}
  {} (\bibinfo {year} {2020}),\ \Eprint {http://arxiv.org/abs/2011.01268}
  {arXiv:2011.01268 [cond-mat.str-el]} \BibitemShut {NoStop}%
\bibitem [{\citenamefont {Ritter}\ \emph {et~al.}(2022)\citenamefont {Ritter},
  \citenamefont {Kiese}, \citenamefont {M{\"u}ller}, \citenamefont {Kugler},
  \citenamefont {Thomale}, \citenamefont {Trebst},\ and\ \citenamefont {von
  Delft}}]{ritter22}%
  \BibitemOpen
  \bibfield  {author} {\bibinfo {author} {\bibfnamefont {M.~K.}\ \bibnamefont
  {Ritter}}, \bibinfo {author} {\bibfnamefont {D.}~\bibnamefont {Kiese}},
  \bibinfo {author} {\bibfnamefont {T.}~\bibnamefont {M{\"u}ller}}, \bibinfo
  {author} {\bibfnamefont {F.~B.}\ \bibnamefont {Kugler}}, \bibinfo {author}
  {\bibfnamefont {R.}~\bibnamefont {Thomale}}, \bibinfo {author} {\bibfnamefont
  {S.}~\bibnamefont {Trebst}}, \ and\ \bibinfo {author} {\bibfnamefont
  {J.}~\bibnamefont {von Delft}},\ }\href {\doibase
  10.1140/epjb/s10051-022-00349-2} {\bibfield  {journal} {\bibinfo  {journal}
  {The European Physical Journal B}\ }\textbf {\bibinfo {volume} {95}},\
  \bibinfo {pages} {102} (\bibinfo {year} {2022})}\BibitemShut {NoStop}%
\bibitem [{\citenamefont {Kiese}\ \emph {et~al.}(2022)\citenamefont {Kiese},
  \citenamefont {M\"uller}, \citenamefont {Iqbal}, \citenamefont {Thomale},\
  and\ \citenamefont {Trebst}}]{kiese22}%
  \BibitemOpen
  \bibfield  {author} {\bibinfo {author} {\bibfnamefont {D.}~\bibnamefont
  {Kiese}}, \bibinfo {author} {\bibfnamefont {T.}~\bibnamefont {M\"uller}},
  \bibinfo {author} {\bibfnamefont {Y.}~\bibnamefont {Iqbal}}, \bibinfo
  {author} {\bibfnamefont {R.}~\bibnamefont {Thomale}}, \ and\ \bibinfo
  {author} {\bibfnamefont {S.}~\bibnamefont {Trebst}},\ }\href {\doibase
  10.1103/PhysRevResearch.4.023185} {\bibfield  {journal} {\bibinfo  {journal}
  {Phys. Rev. Research}\ }\textbf {\bibinfo {volume} {4}},\ \bibinfo {pages}
  {023185} (\bibinfo {year} {2022})}\BibitemShut {NoStop}%
\bibitem [{\citenamefont {Schneider}\ \emph {et~al.}(2022)\citenamefont
  {Schneider}, \citenamefont {Kiese},\ and\ \citenamefont
  {Sbierski}}]{schneider22}%
  \BibitemOpen
  \bibfield  {author} {\bibinfo {author} {\bibfnamefont {B.}~\bibnamefont
  {Schneider}}, \bibinfo {author} {\bibfnamefont {D.}~\bibnamefont {Kiese}}, \
  and\ \bibinfo {author} {\bibfnamefont {B.}~\bibnamefont {Sbierski}},\
  }\href@noop {} {} (\bibinfo {year} {2022}),\ \Eprint
  {http://arxiv.org/abs/2209.13484} {arXiv:2209.13484 [cond-mat.str-el]}
  \BibitemShut {NoStop}%
\bibitem [{\citenamefont {Arovas}\ and\ \citenamefont
  {Auerbach}(1988)}]{arovas88}%
  \BibitemOpen
  \bibfield  {author} {\bibinfo {author} {\bibfnamefont {D.~P.}\ \bibnamefont
  {Arovas}}\ and\ \bibinfo {author} {\bibfnamefont {A.}~\bibnamefont
  {Auerbach}},\ }\href {\doibase 10.1103/PhysRevB.38.316} {\bibfield  {journal}
  {\bibinfo  {journal} {Phys. Rev. B}\ }\textbf {\bibinfo {volume} {38}},\
  \bibinfo {pages} {316} (\bibinfo {year} {1988})}\BibitemShut {NoStop}%
\bibitem [{\citenamefont {Auerbach}(1994)}]{auerbach_book}%
  \BibitemOpen
  \bibfield  {author} {\bibinfo {author} {\bibfnamefont {A.}~\bibnamefont
  {Auerbach}},\ }\href@noop {} {\emph {\bibinfo {title} {Interacting Electrons
  and Quantum Magnetism}}}\ (\bibinfo  {publisher} {Springer},\ \bibinfo {year}
  {1994})\BibitemShut {NoStop}%
\bibitem [{\citenamefont {Zhang}\ \emph {et~al.}(2022)\citenamefont {Zhang},
  \citenamefont {Ghioldi}, \citenamefont {Manuel}, \citenamefont {Trumper},\
  and\ \citenamefont {Batista}}]{zhang22}%
  \BibitemOpen
  \bibfield  {author} {\bibinfo {author} {\bibfnamefont {S.-S.}\ \bibnamefont
  {Zhang}}, \bibinfo {author} {\bibfnamefont {E.~A.}\ \bibnamefont {Ghioldi}},
  \bibinfo {author} {\bibfnamefont {L.~O.}\ \bibnamefont {Manuel}}, \bibinfo
  {author} {\bibfnamefont {A.~E.}\ \bibnamefont {Trumper}}, \ and\ \bibinfo
  {author} {\bibfnamefont {C.~D.}\ \bibnamefont {Batista}},\ }\href {\doibase
  10.1103/PhysRevB.105.224404} {\bibfield  {journal} {\bibinfo  {journal}
  {Phys. Rev. B}\ }\textbf {\bibinfo {volume} {105}},\ \bibinfo {pages}
  {224404} (\bibinfo {year} {2022})}\BibitemShut {NoStop}%
\bibitem [{\citenamefont {Hirsch}\ and\ \citenamefont {Tang}(1989)}]{hirsch89}%
  \BibitemOpen
  \bibfield  {author} {\bibinfo {author} {\bibfnamefont {J.~E.}\ \bibnamefont
  {Hirsch}}\ and\ \bibinfo {author} {\bibfnamefont {S.}~\bibnamefont {Tang}},\
  }\href {\doibase 10.1103/PhysRevB.39.2850} {\bibfield  {journal} {\bibinfo
  {journal} {Phys. Rev. B}\ }\textbf {\bibinfo {volume} {39}},\ \bibinfo
  {pages} {2850} (\bibinfo {year} {1989})}\BibitemShut {NoStop}%
\bibitem [{\citenamefont {Sarker}\ \emph {et~al.}(1989)\citenamefont {Sarker},
  \citenamefont {Jayaprakash}, \citenamefont {Krishnamurthy},\ and\
  \citenamefont {Ma}}]{sarker89}%
  \BibitemOpen
  \bibfield  {author} {\bibinfo {author} {\bibfnamefont {S.}~\bibnamefont
  {Sarker}}, \bibinfo {author} {\bibfnamefont {C.}~\bibnamefont {Jayaprakash}},
  \bibinfo {author} {\bibfnamefont {H.~R.}\ \bibnamefont {Krishnamurthy}}, \
  and\ \bibinfo {author} {\bibfnamefont {M.}~\bibnamefont {Ma}},\ }\href
  {\doibase 10.1103/PhysRevB.40.5028} {\bibfield  {journal} {\bibinfo
  {journal} {Phys. Rev. B}\ }\textbf {\bibinfo {volume} {40}},\ \bibinfo
  {pages} {5028} (\bibinfo {year} {1989})}\BibitemShut {NoStop}%
\bibitem [{\citenamefont {Chandra}\ \emph {et~al.}(1990)\citenamefont
  {Chandra}, \citenamefont {Coleman},\ and\ \citenamefont
  {Larkin}}]{chandra90}%
  \BibitemOpen
  \bibfield  {author} {\bibinfo {author} {\bibfnamefont {P.}~\bibnamefont
  {Chandra}}, \bibinfo {author} {\bibfnamefont {P.}~\bibnamefont {Coleman}}, \
  and\ \bibinfo {author} {\bibfnamefont {A.~I.}\ \bibnamefont {Larkin}},\
  }\href {\doibase 10.1088/0953-8984/2/39/008} {\bibfield  {journal} {\bibinfo
  {journal} {Journal of Physics: Condensed Matter}\ }\textbf {\bibinfo {volume}
  {2}},\ \bibinfo {pages} {7933} (\bibinfo {year} {1990})}\BibitemShut
  {NoStop}%
\bibitem [{\citenamefont {Kitaev}(2006)}]{kitaev_anyons_2006}%
  \BibitemOpen
  \bibfield  {author} {\bibinfo {author} {\bibfnamefont {A.}~\bibnamefont
  {Kitaev}},\ }\href {\doibase 10.1016/j.aop.2005.10.005} {\bibfield  {journal}
  {\bibinfo  {journal} {Annals of Physics}\ }\textbf {\bibinfo {volume}
  {321}},\ \bibinfo {pages} {2} (\bibinfo {year} {2006})}\BibitemShut {NoStop}%
\bibitem [{\citenamefont {Yao}\ and\ \citenamefont {Kivelson}(2007)}]{yao07}%
  \BibitemOpen
  \bibfield  {author} {\bibinfo {author} {\bibfnamefont {H.}~\bibnamefont
  {Yao}}\ and\ \bibinfo {author} {\bibfnamefont {S.~A.}\ \bibnamefont
  {Kivelson}},\ }\href {\doibase 10.1103/PhysRevLett.99.247203} {\bibfield
  {journal} {\bibinfo  {journal} {Phys. Rev. Lett.}\ }\textbf {\bibinfo
  {volume} {99}},\ \bibinfo {pages} {247203} (\bibinfo {year}
  {2007})}\BibitemShut {NoStop}%
\bibitem [{\citenamefont {Mandal}\ and\ \citenamefont
  {Surendran}(2009)}]{mandal09}%
  \BibitemOpen
  \bibfield  {author} {\bibinfo {author} {\bibfnamefont {S.}~\bibnamefont
  {Mandal}}\ and\ \bibinfo {author} {\bibfnamefont {N.}~\bibnamefont
  {Surendran}},\ }\href {\doibase 10.1103/PhysRevB.79.024426} {\bibfield
  {journal} {\bibinfo  {journal} {Phys. Rev. B}\ }\textbf {\bibinfo {volume}
  {79}},\ \bibinfo {pages} {024426} (\bibinfo {year} {2009})}\BibitemShut
  {NoStop}%
\bibitem [{\citenamefont {Hermanns}\ and\ \citenamefont
  {Trebst}(2014)}]{hermanns14}%
  \BibitemOpen
  \bibfield  {author} {\bibinfo {author} {\bibfnamefont {M.}~\bibnamefont
  {Hermanns}}\ and\ \bibinfo {author} {\bibfnamefont {S.}~\bibnamefont
  {Trebst}},\ }\href {\doibase 10.1103/PhysRevB.89.235102} {\bibfield
  {journal} {\bibinfo  {journal} {Phys. Rev. B}\ }\textbf {\bibinfo {volume}
  {89}},\ \bibinfo {pages} {235102} (\bibinfo {year} {2014})}\BibitemShut
  {NoStop}%
\bibitem [{\citenamefont {Hermanns}\ \emph {et~al.}(2015)\citenamefont
  {Hermanns}, \citenamefont {O'Brien},\ and\ \citenamefont
  {Trebst}}]{hermanns15}%
  \BibitemOpen
  \bibfield  {author} {\bibinfo {author} {\bibfnamefont {M.}~\bibnamefont
  {Hermanns}}, \bibinfo {author} {\bibfnamefont {K.}~\bibnamefont {O'Brien}}, \
  and\ \bibinfo {author} {\bibfnamefont {S.}~\bibnamefont {Trebst}},\ }\href
  {\doibase 10.1103/PhysRevLett.114.157202} {\bibfield  {journal} {\bibinfo
  {journal} {Phys. Rev. Lett.}\ }\textbf {\bibinfo {volume} {114}},\ \bibinfo
  {pages} {157202} (\bibinfo {year} {2015})}\BibitemShut {NoStop}%
\bibitem [{\citenamefont {O'Brien}\ \emph {et~al.}(2016)\citenamefont
  {O'Brien}, \citenamefont {Hermanns},\ and\ \citenamefont
  {Trebst}}]{obrien16}%
  \BibitemOpen
  \bibfield  {author} {\bibinfo {author} {\bibfnamefont {K.}~\bibnamefont
  {O'Brien}}, \bibinfo {author} {\bibfnamefont {M.}~\bibnamefont {Hermanns}}, \
  and\ \bibinfo {author} {\bibfnamefont {S.}~\bibnamefont {Trebst}},\ }\href
  {\doibase 10.1103/PhysRevB.93.085101} {\bibfield  {journal} {\bibinfo
  {journal} {Phys. Rev. B}\ }\textbf {\bibinfo {volume} {93}},\ \bibinfo
  {pages} {085101} (\bibinfo {year} {2016})}\BibitemShut {NoStop}%
\bibitem [{\citenamefont {Rachel}\ \emph {et~al.}(2016)\citenamefont {Rachel},
  \citenamefont {Fritz},\ and\ \citenamefont {Vojta}}]{rachel16}%
  \BibitemOpen
  \bibfield  {author} {\bibinfo {author} {\bibfnamefont {S.}~\bibnamefont
  {Rachel}}, \bibinfo {author} {\bibfnamefont {L.}~\bibnamefont {Fritz}}, \
  and\ \bibinfo {author} {\bibfnamefont {M.}~\bibnamefont {Vojta}},\ }\href
  {\doibase 10.1103/PhysRevLett.116.167201} {\bibfield  {journal} {\bibinfo
  {journal} {Phys. Rev. Lett.}\ }\textbf {\bibinfo {volume} {116}},\ \bibinfo
  {pages} {167201} (\bibinfo {year} {2016})}\BibitemShut {NoStop}%
\bibitem [{\citenamefont {Cassella}\ \emph {et~al.}(2022)\citenamefont
  {Cassella}, \citenamefont {D'Ornellas}, \citenamefont {Hodson}, \citenamefont
  {Natori},\ and\ \citenamefont {Knolle}}]{casella22}%
  \BibitemOpen
  \bibfield  {author} {\bibinfo {author} {\bibfnamefont {G.}~\bibnamefont
  {Cassella}}, \bibinfo {author} {\bibfnamefont {P.}~\bibnamefont
  {D'Ornellas}}, \bibinfo {author} {\bibfnamefont {T.}~\bibnamefont {Hodson}},
  \bibinfo {author} {\bibfnamefont {W.~M.~H.}\ \bibnamefont {Natori}}, \ and\
  \bibinfo {author} {\bibfnamefont {J.}~\bibnamefont {Knolle}},\ }\href@noop {}
  {} (\bibinfo {year} {2022}),\ \Eprint {http://arxiv.org/abs/2208.08246}
  {arXiv:2208.08246 [cond-mat.str-el]} \BibitemShut {NoStop}%
\bibitem [{\citenamefont {Martin}(1959)}]{martin59}%
  \BibitemOpen
  \bibfield  {author} {\bibinfo {author} {\bibfnamefont {J.~L.}\ \bibnamefont
  {Martin}},\ }\href {\doibase 10.1098/rspa.1959.0126} {\bibfield  {journal}
  {\bibinfo  {journal} {Proc. R. Soc. Lond. A}\ }\textbf {\bibinfo {volume}
  {251}},\ \bibinfo {pages} {536} (\bibinfo {year} {1959})}\BibitemShut
  {NoStop}%
\bibitem [{\citenamefont {Tsvelik}(1992)}]{tsvelik92}%
  \BibitemOpen
  \bibfield  {author} {\bibinfo {author} {\bibfnamefont {A.~M.}\ \bibnamefont
  {Tsvelik}},\ }\href {\doibase 10.1103/PhysRevLett.69.2142} {\bibfield
  {journal} {\bibinfo  {journal} {Phys. Rev. Lett.}\ }\textbf {\bibinfo
  {volume} {69}},\ \bibinfo {pages} {2142} (\bibinfo {year}
  {1992})}\BibitemShut {NoStop}%
\bibitem [{\citenamefont {Chen}\ \emph {et~al.}(2012)\citenamefont {Chen},
  \citenamefont {Essin},\ and\ \citenamefont {Hermele}}]{chen12}%
  \BibitemOpen
  \bibfield  {author} {\bibinfo {author} {\bibfnamefont {G.}~\bibnamefont
  {Chen}}, \bibinfo {author} {\bibfnamefont {A.}~\bibnamefont {Essin}}, \ and\
  \bibinfo {author} {\bibfnamefont {M.}~\bibnamefont {Hermele}},\ }\href
  {\doibase 10.1103/PhysRevB.85.094418} {\bibfield  {journal} {\bibinfo
  {journal} {Phys. Rev. B}\ }\textbf {\bibinfo {volume} {85}},\ \bibinfo
  {pages} {094418} (\bibinfo {year} {2012})}\BibitemShut {NoStop}%
\bibitem [{\citenamefont {Shastry}\ and\ \citenamefont
  {Sen}(1997)}]{shastry97}%
  \BibitemOpen
  \bibfield  {author} {\bibinfo {author} {\bibfnamefont {B.~S.}\ \bibnamefont
  {Shastry}}\ and\ \bibinfo {author} {\bibfnamefont {D.}~\bibnamefont {Sen}},\
  }\href {\doibase 10.1103/PhysRevB.55.2988} {\bibfield  {journal} {\bibinfo
  {journal} {Phys. Rev. B}\ }\textbf {\bibinfo {volume} {55}},\ \bibinfo
  {pages} {2988} (\bibinfo {year} {1997})}\BibitemShut {NoStop}%
\bibitem [{\citenamefont {Mao}\ \emph {et~al.}(2003)\citenamefont {Mao},
  \citenamefont {Coleman}, \citenamefont {Hooley},\ and\ \citenamefont
  {Langreth}}]{mao03}%
  \BibitemOpen
  \bibfield  {author} {\bibinfo {author} {\bibfnamefont {W.}~\bibnamefont
  {Mao}}, \bibinfo {author} {\bibfnamefont {P.}~\bibnamefont {Coleman}},
  \bibinfo {author} {\bibfnamefont {C.}~\bibnamefont {Hooley}}, \ and\ \bibinfo
  {author} {\bibfnamefont {D.}~\bibnamefont {Langreth}},\ }\href {\doibase
  10.1103/PhysRevLett.91.207203} {\bibfield  {journal} {\bibinfo  {journal}
  {Phys. Rev. Lett.}\ }\textbf {\bibinfo {volume} {91}},\ \bibinfo {pages}
  {207203} (\bibinfo {year} {2003})}\BibitemShut {NoStop}%
\bibitem [{\citenamefont {Shnirman}\ and\ \citenamefont
  {Makhlin}(2003)}]{shnirman03}%
  \BibitemOpen
  \bibfield  {author} {\bibinfo {author} {\bibfnamefont {A.}~\bibnamefont
  {Shnirman}}\ and\ \bibinfo {author} {\bibfnamefont {Y.}~\bibnamefont
  {Makhlin}},\ }\href {\doibase 10.1103/PhysRevLett.91.207204} {\bibfield
  {journal} {\bibinfo  {journal} {Phys. Rev. Lett.}\ }\textbf {\bibinfo
  {volume} {91}},\ \bibinfo {pages} {207204} (\bibinfo {year}
  {2003})}\BibitemShut {NoStop}%
\bibitem [{\citenamefont {Biswas}\ \emph {et~al.}(2011)\citenamefont {Biswas},
  \citenamefont {Fu}, \citenamefont {Laumann},\ and\ \citenamefont
  {Sachdev}}]{biswas_su2-invariant_2011}%
  \BibitemOpen
  \bibfield  {author} {\bibinfo {author} {\bibfnamefont {R.~R.}\ \bibnamefont
  {Biswas}}, \bibinfo {author} {\bibfnamefont {L.}~\bibnamefont {Fu}}, \bibinfo
  {author} {\bibfnamefont {C.~R.}\ \bibnamefont {Laumann}}, \ and\ \bibinfo
  {author} {\bibfnamefont {S.}~\bibnamefont {Sachdev}},\ }\href {\doibase
  10.1103/PhysRevB.83.245131} {\bibfield  {journal} {\bibinfo  {journal} {Phys.
  Rev. B}\ }\textbf {\bibinfo {volume} {83}},\ \bibinfo {pages} {245131}
  (\bibinfo {year} {2011})}\BibitemShut {NoStop}%
\bibitem [{\citenamefont {Schad}\ \emph {et~al.}(2014)\citenamefont {Schad},
  \citenamefont {Narozhny}, \citenamefont {Sch\"on},\ and\ \citenamefont
  {Shnirman}}]{schad14}%
  \BibitemOpen
  \bibfield  {author} {\bibinfo {author} {\bibfnamefont {P.}~\bibnamefont
  {Schad}}, \bibinfo {author} {\bibfnamefont {B.~N.}\ \bibnamefont {Narozhny}},
  \bibinfo {author} {\bibfnamefont {G.}~\bibnamefont {Sch\"on}}, \ and\
  \bibinfo {author} {\bibfnamefont {A.}~\bibnamefont {Shnirman}},\ }\href
  {\doibase 10.1103/PhysRevB.90.205419} {\bibfield  {journal} {\bibinfo
  {journal} {Phys. Rev. B}\ }\textbf {\bibinfo {volume} {90}},\ \bibinfo
  {pages} {205419} (\bibinfo {year} {2014})}\BibitemShut {NoStop}%
\bibitem [{\citenamefont {Schad}\ \emph {et~al.}(2015)\citenamefont {Schad},
  \citenamefont {Makhlin}, \citenamefont {Narozhny}, \citenamefont {Sch\"on},\
  and\ \citenamefont {Shnirman}}]{schad15}%
  \BibitemOpen
  \bibfield  {author} {\bibinfo {author} {\bibfnamefont {P.}~\bibnamefont
  {Schad}}, \bibinfo {author} {\bibfnamefont {Y.}~\bibnamefont {Makhlin}},
  \bibinfo {author} {\bibfnamefont {B.}~\bibnamefont {Narozhny}}, \bibinfo
  {author} {\bibfnamefont {G.}~\bibnamefont {Sch\"on}}, \ and\ \bibinfo {author}
  {\bibfnamefont {A.}~\bibnamefont {Shnirman}},\ }\href {\doibase
  https://doi.org/10.1016/j.aop.2015.07.006} {\bibfield  {journal} {\bibinfo
  {journal} {Annals of Physics}\ }\textbf {\bibinfo {volume} {361}},\ \bibinfo
  {pages} {401} (\bibinfo {year} {2015})}\BibitemShut {NoStop}%
\bibitem [{\citenamefont {Schad}\ \emph {et~al.}(2016)\citenamefont {Schad},
  \citenamefont {Shnirman},\ and\ \citenamefont {Makhlin}}]{schad16}%
  \BibitemOpen
  \bibfield  {author} {\bibinfo {author} {\bibfnamefont {P.}~\bibnamefont
  {Schad}}, \bibinfo {author} {\bibfnamefont {A.}~\bibnamefont {Shnirman}}, \
  and\ \bibinfo {author} {\bibfnamefont {Y.}~\bibnamefont {Makhlin}},\ }\href
  {\doibase 10.1103/PhysRevB.93.174420} {\bibfield  {journal} {\bibinfo
  {journal} {Phys. Rev. B}\ }\textbf {\bibinfo {volume} {93}},\ \bibinfo
  {pages} {174420} (\bibinfo {year} {2016})}\BibitemShut {NoStop}%
\bibitem [{\citenamefont {Niggemann}\ \emph {et~al.}(2021)\citenamefont
  {Niggemann}, \citenamefont {Sbierski},\ and\ \citenamefont
  {Reuther}}]{niggemann_frustrated_2021}%
  \BibitemOpen
  \bibfield  {author} {\bibinfo {author} {\bibfnamefont {N.}~\bibnamefont
  {Niggemann}}, \bibinfo {author} {\bibfnamefont {B.}~\bibnamefont {Sbierski}},
  \ and\ \bibinfo {author} {\bibfnamefont {J.}~\bibnamefont {Reuther}},\ }\href
  {\doibase 10.1103/PhysRevB.103.104431} {\bibfield  {journal} {\bibinfo
  {journal} {Phys. Rev. B}\ }\textbf {\bibinfo {volume} {103}},\ \bibinfo
  {pages} {104431} (\bibinfo {year} {2021})}\BibitemShut {NoStop}%
\bibitem [{\citenamefont {Niggemann}\ \emph {et~al.}(2022)\citenamefont
  {Niggemann}, \citenamefont {Reuther},\ and\ \citenamefont
  {Sbierski}}]{niggemann_quantitative_2022}%
  \BibitemOpen
  \bibfield  {author} {\bibinfo {author} {\bibfnamefont {N.}~\bibnamefont
  {Niggemann}}, \bibinfo {author} {\bibfnamefont {J.}~\bibnamefont {Reuther}},
  \ and\ \bibinfo {author} {\bibfnamefont {B.}~\bibnamefont {Sbierski}},\
  }\href {\doibase 10.21468/SciPostPhys.12.5.156} {\bibfield  {journal}
  {\bibinfo  {journal} {SciPost Phys.}\ }\textbf {\bibinfo {volume} {12}},\
  \bibinfo {pages} {156} (\bibinfo {year} {2022})}\BibitemShut {NoStop}%
\bibitem [{\citenamefont {Wang}\ and\ \citenamefont
  {Vishwanath}(2009)}]{wang_z_2009}%
  \BibitemOpen
  \bibfield  {author} {\bibinfo {author} {\bibfnamefont {F.}~\bibnamefont
  {Wang}}\ and\ \bibinfo {author} {\bibfnamefont {A.}~\bibnamefont
  {Vishwanath}},\ }\href {\doibase 10.1103/PhysRevB.80.064413} {\bibfield
  {journal} {\bibinfo  {journal} {Phys. Rev. B}\ }\textbf {\bibinfo {volume}
  {80}},\ \bibinfo {pages} {064413} (\bibinfo {year} {2009})}\BibitemShut
  {NoStop}%
\bibitem [{\citenamefont {Yao}\ \emph {et~al.}(2009)\citenamefont {Yao},
  \citenamefont {Zhang},\ and\ \citenamefont {Kivelson}}]{yao09}%
  \BibitemOpen
  \bibfield  {author} {\bibinfo {author} {\bibfnamefont {H.}~\bibnamefont
  {Yao}}, \bibinfo {author} {\bibfnamefont {S.-C.}\ \bibnamefont {Zhang}}, \
  and\ \bibinfo {author} {\bibfnamefont {S.~A.}\ \bibnamefont {Kivelson}},\
  }\href {\doibase 10.1103/PhysRevLett.102.217202} {\bibfield  {journal}
  {\bibinfo  {journal} {Phys. Rev. Lett.}\ }\textbf {\bibinfo {volume} {102}},\
  \bibinfo {pages} {217202} (\bibinfo {year} {2009})}\BibitemShut {NoStop}%
\bibitem [{\citenamefont {Yao}\ and\ \citenamefont {Lee}(2011)}]{yao11}%
  \BibitemOpen
  \bibfield  {author} {\bibinfo {author} {\bibfnamefont {H.}~\bibnamefont
  {Yao}}\ and\ \bibinfo {author} {\bibfnamefont {D.-H.}\ \bibnamefont {Lee}},\
  }\href {\doibase 10.1103/PhysRevLett.107.087205} {\bibfield  {journal}
  {\bibinfo  {journal} {Phys. Rev. Lett.}\ }\textbf {\bibinfo {volume} {107}},\
  \bibinfo {pages} {087205} (\bibinfo {year} {2011})}\BibitemShut {NoStop}%
\bibitem [{\citenamefont {Chua}\ \emph {et~al.}(2011)\citenamefont {Chua},
  \citenamefont {Yao},\ and\ \citenamefont {Fiete}}]{chua11}%
  \BibitemOpen
  \bibfield  {author} {\bibinfo {author} {\bibfnamefont {V.}~\bibnamefont
  {Chua}}, \bibinfo {author} {\bibfnamefont {H.}~\bibnamefont {Yao}}, \ and\
  \bibinfo {author} {\bibfnamefont {G.~A.}\ \bibnamefont {Fiete}},\ }\href
  {\doibase 10.1103/PhysRevB.83.180412} {\bibfield  {journal} {\bibinfo
  {journal} {Phys. Rev. B}\ }\textbf {\bibinfo {volume} {83}},\ \bibinfo
  {pages} {180412} (\bibinfo {year} {2011})}\BibitemShut {NoStop}%
\bibitem [{\citenamefont {Natori}\ \emph {et~al.}(2016)\citenamefont {Natori},
  \citenamefont {Andrade}, \citenamefont {Miranda},\ and\ \citenamefont
  {Pereira}}]{natori_chiral_2016}%
  \BibitemOpen
  \bibfield  {author} {\bibinfo {author} {\bibfnamefont {W.}~\bibnamefont
  {Natori}}, \bibinfo {author} {\bibfnamefont {E.}~\bibnamefont {Andrade}},
  \bibinfo {author} {\bibfnamefont {E.}~\bibnamefont {Miranda}}, \ and\
  \bibinfo {author} {\bibfnamefont {R.}~\bibnamefont {Pereira}},\ }\href
  {\doibase 10.1103/PhysRevLett.117.017204} {\bibfield  {journal} {\bibinfo
  {journal} {Phys. Rev. Lett.}\ }\textbf {\bibinfo {volume} {117}},\ \bibinfo
  {pages} {017204} (\bibinfo {year} {2016})}\BibitemShut {NoStop}%
\bibitem [{\citenamefont {Natori}\ \emph {et~al.}(2017)\citenamefont {Natori},
  \citenamefont {Daghofer},\ and\ \citenamefont
  {Pereira}}]{natori_dynamics_2017}%
  \BibitemOpen
  \bibfield  {author} {\bibinfo {author} {\bibfnamefont {W.~M.~H.}\
  \bibnamefont {Natori}}, \bibinfo {author} {\bibfnamefont {M.}~\bibnamefont
  {Daghofer}}, \ and\ \bibinfo {author} {\bibfnamefont {R.~G.}\ \bibnamefont
  {Pereira}},\ }\href {\doibase 10.1103/PhysRevB.96.125109} {\bibfield
  {journal} {\bibinfo  {journal} {Phys. Rev. B}\ }\textbf {\bibinfo {volume}
  {96}},\ \bibinfo {pages} {125109} (\bibinfo {year} {2017})}\BibitemShut
  {NoStop}%
\bibitem [{\citenamefont {Natori}\ \emph {et~al.}(2018)\citenamefont {Natori},
  \citenamefont {Andrade},\ and\ \citenamefont {Pereira}}]{natori18}%
  \BibitemOpen
  \bibfield  {author} {\bibinfo {author} {\bibfnamefont {W.~M.~H.}\
  \bibnamefont {Natori}}, \bibinfo {author} {\bibfnamefont {E.~C.}\
  \bibnamefont {Andrade}}, \ and\ \bibinfo {author} {\bibfnamefont {R.~G.}\
  \bibnamefont {Pereira}},\ }\href {\doibase 10.1103/PhysRevB.98.195113}
  {\bibfield  {journal} {\bibinfo  {journal} {Phys. Rev. B}\ }\textbf {\bibinfo
  {volume} {98}},\ \bibinfo {pages} {195113} (\bibinfo {year}
  {2018})}\BibitemShut {NoStop}%
\bibitem [{\citenamefont {de~Carvalho}\ \emph {et~al.}(2018)\citenamefont
  {de~Carvalho}, \citenamefont {Freire}, \citenamefont {Miranda},\ and\
  \citenamefont {Pereira}}]{carvalho18}%
  \BibitemOpen
  \bibfield  {author} {\bibinfo {author} {\bibfnamefont {V.~S.}\ \bibnamefont
  {de~Carvalho}}, \bibinfo {author} {\bibfnamefont {H.}~\bibnamefont {Freire}},
  \bibinfo {author} {\bibfnamefont {E.}~\bibnamefont {Miranda}}, \ and\
  \bibinfo {author} {\bibfnamefont {R.~G.}\ \bibnamefont {Pereira}},\ }\href
  {\doibase 10.1103/PhysRevB.98.155105} {\bibfield  {journal} {\bibinfo
  {journal} {Phys. Rev. B}\ }\textbf {\bibinfo {volume} {98}},\ \bibinfo
  {pages} {155105} (\bibinfo {year} {2018})}\BibitemShut {NoStop}%
\bibitem [{\citenamefont {de~Farias}\ \emph {et~al.}(2020)\citenamefont
  {de~Farias}, \citenamefont {de~Carvalho}, \citenamefont {Miranda},\ and\
  \citenamefont {Pereira}}]{de_farias_quadrupolar_2020}%
  \BibitemOpen
  \bibfield  {author} {\bibinfo {author} {\bibfnamefont {C.~S.}\ \bibnamefont
  {de~Farias}}, \bibinfo {author} {\bibfnamefont {V.~S.}\ \bibnamefont
  {de~Carvalho}}, \bibinfo {author} {\bibfnamefont {E.}~\bibnamefont
  {Miranda}}, \ and\ \bibinfo {author} {\bibfnamefont {R.~G.}\ \bibnamefont
  {Pereira}},\ }\href {\doibase 10.1103/PhysRevB.102.075110} {\bibfield
  {journal} {\bibinfo  {journal} {Phys. Rev. B}\ }\textbf {\bibinfo {volume}
  {102}},\ \bibinfo {pages} {075110} (\bibinfo {year} {2020})}\BibitemShut
  {NoStop}%
\bibitem [{\citenamefont {Natori}\ and\ \citenamefont
  {Knolle}(2020)}]{natori20}%
  \BibitemOpen
  \bibfield  {author} {\bibinfo {author} {\bibfnamefont {W.~M.~H.}\
  \bibnamefont {Natori}}\ and\ \bibinfo {author} {\bibfnamefont
  {J.}~\bibnamefont {Knolle}},\ }\href {\doibase
  10.1103/PhysRevLett.125.067201} {\bibfield  {journal} {\bibinfo  {journal}
  {Phys. Rev. Lett.}\ }\textbf {\bibinfo {volume} {125}},\ \bibinfo {pages}
  {067201} (\bibinfo {year} {2020})}\BibitemShut {NoStop}%
\bibitem [{\citenamefont {Jin}\ \emph {et~al.}(2022)\citenamefont {Jin},
  \citenamefont {Natori}, \citenamefont {Pollmann},\ and\ \citenamefont
  {Knolle}}]{jin_unveiling_2022}%
  \BibitemOpen
  \bibfield  {author} {\bibinfo {author} {\bibfnamefont {H.-K.}\ \bibnamefont
  {Jin}}, \bibinfo {author} {\bibfnamefont {W.~M.~H.}\ \bibnamefont {Natori}},
  \bibinfo {author} {\bibfnamefont {F.}~\bibnamefont {Pollmann}}, \ and\
  \bibinfo {author} {\bibfnamefont {J.}~\bibnamefont {Knolle}},\ }\href
  {\doibase 10.1038/s41467-022-31503-0} {\bibfield  {journal} {\bibinfo
  {journal} {Nat Commun}\ }\textbf {\bibinfo {volume} {13}},\ \bibinfo {pages}
  {3813} (\bibinfo {year} {2022})}\BibitemShut {NoStop}%
\bibitem [{\citenamefont {Alet}\ \emph {et~al.}(2011)\citenamefont {Alet},
  \citenamefont {Capponi}, \citenamefont {Nonne}, \citenamefont {Lecheminant},\
  and\ \citenamefont {McCulloch}}]{PhysRevB.83.060407}%
  \BibitemOpen
  \bibfield  {author} {\bibinfo {author} {\bibfnamefont {F.}~\bibnamefont
  {Alet}}, \bibinfo {author} {\bibfnamefont {S.}~\bibnamefont {Capponi}},
  \bibinfo {author} {\bibfnamefont {H.}~\bibnamefont {Nonne}}, \bibinfo
  {author} {\bibfnamefont {P.}~\bibnamefont {Lecheminant}}, \ and\ \bibinfo
  {author} {\bibfnamefont {I.~P.}\ \bibnamefont {McCulloch}},\ }\href {\doibase
  10.1103/PhysRevB.83.060407} {\bibfield  {journal} {\bibinfo  {journal} {Phys.
  Rev. B}\ }\textbf {\bibinfo {volume} {83}},\ \bibinfo {pages} {060407}
  (\bibinfo {year} {2011})}\BibitemShut {NoStop}%
\bibitem [{\citenamefont {Verresen}\ and\ \citenamefont
  {Vishwanath}(2022)}]{PhysRevX.12.041029}%
  \BibitemOpen
  \bibfield  {author} {\bibinfo {author} {\bibfnamefont {R.}~\bibnamefont
  {Verresen}}\ and\ \bibinfo {author} {\bibfnamefont {A.}~\bibnamefont
  {Vishwanath}},\ }\href {\doibase 10.1103/PhysRevX.12.041029} {\bibfield
  {journal} {\bibinfo  {journal} {Phys. Rev. X}\ }\textbf {\bibinfo {volume}
  {12}},\ \bibinfo {pages} {041029} (\bibinfo {year} {2022})}\BibitemShut
  {NoStop}%
\bibitem [{\citenamefont {Zee}(2016)}]{group_theory}%
  \BibitemOpen
  \bibfield  {author} {\bibinfo {author} {\bibfnamefont {A.}~\bibnamefont
  {Zee}},\ }\href@noop {} {\emph {\bibinfo {title} {Group Theory in a Nutshell
  for Physicists}}}\ (\bibinfo  {publisher} {Princeton University Press},\
  \bibinfo {year} {2016})\BibitemShut {NoStop}%
\bibitem [{\citenamefont {Popov}\ and\ \citenamefont
  {Fedotov}(1988)}]{popov88}%
  \BibitemOpen
  \bibfield  {author} {\bibinfo {author} {\bibfnamefont {V.~N.}\ \bibnamefont
  {Popov}}\ and\ \bibinfo {author} {\bibfnamefont {S.~A.}\ \bibnamefont
  {Fedotov}},\ }\href
  {http://www.jetp.ras.ru/cgi-bin/e/index/e/67/3/p535?a=list} {\bibfield
  {journal} {\bibinfo  {journal} {Sov. Phys. JETP}\ }\textbf {\bibinfo {volume}
  {67}},\ \bibinfo {pages} {535} (\bibinfo {year} {1988})}\BibitemShut
  {NoStop}%
\bibitem [{\citenamefont {Prokof'ev}\ and\ \citenamefont
  {Svistunov}(2011)}]{prokofiev11}%
  \BibitemOpen
  \bibfield  {author} {\bibinfo {author} {\bibfnamefont {N.~V.}\ \bibnamefont
  {Prokof'ev}}\ and\ \bibinfo {author} {\bibfnamefont {B.~V.}\ \bibnamefont
  {Svistunov}},\ }\href {\doibase 10.1103/PhysRevB.84.073102} {\bibfield
  {journal} {\bibinfo  {journal} {Phys. Rev. B}\ }\textbf {\bibinfo {volume}
  {84}},\ \bibinfo {pages} {073102} (\bibinfo {year} {2011})}\BibitemShut
  {NoStop}%
\bibitem [{\citenamefont {Itzkowitz}\ \emph {et~al.}(1991)\citenamefont
  {Itzkowitz}, \citenamefont {Rothman},\ and\ \citenamefont
  {Strassberg}}]{itzkowitz_note_1991}%
  \BibitemOpen
  \bibfield  {author} {\bibinfo {author} {\bibfnamefont {G.}~\bibnamefont
  {Itzkowitz}}, \bibinfo {author} {\bibfnamefont {S.}~\bibnamefont {Rothman}},
  \ and\ \bibinfo {author} {\bibfnamefont {H.}~\bibnamefont {Strassberg}},\
  }\href {\doibase 10.1016/0022-4049(91)90023-U} {\bibfield  {journal}
  {\bibinfo  {journal} {Journal of Pure and Applied Algebra}\ }\textbf
  {\bibinfo {volume} {69}},\ \bibinfo {pages} {285} (\bibinfo {year}
  {1991})}\BibitemShut {NoStop}%
\bibitem [{\citenamefont {Wen}(1999)}]{wen99}%
  \BibitemOpen
  \bibfield  {author} {\bibinfo {author} {\bibfnamefont {X.-G.}\ \bibnamefont
  {Wen}},\ }\href {\doibase 10.1103/PhysRevB.60.8827} {\bibfield  {journal}
  {\bibinfo  {journal} {Phys. Rev. B}\ }\textbf {\bibinfo {volume} {60}},\
  \bibinfo {pages} {8827} (\bibinfo {year} {1999})}\BibitemShut {NoStop}%
\bibitem [{\citenamefont {Barkeshli}\ and\ \citenamefont
  {Wen}(2010)}]{barkeshli10}%
  \BibitemOpen
  \bibfield  {author} {\bibinfo {author} {\bibfnamefont {M.}~\bibnamefont
  {Barkeshli}}\ and\ \bibinfo {author} {\bibfnamefont {X.-G.}\ \bibnamefont
  {Wen}},\ }\href {\doibase 10.1103/PhysRevB.81.155302} {\bibfield  {journal}
  {\bibinfo  {journal} {Phys. Rev. B}\ }\textbf {\bibinfo {volume} {81}},\
  \bibinfo {pages} {155302} (\bibinfo {year} {2010})}\BibitemShut {NoStop}%
\bibitem [{\citenamefont {Ma}(2022)}]{ma22}%
  \BibitemOpen
  \bibfield  {author} {\bibinfo {author} {\bibfnamefont {H.}~\bibnamefont
  {Ma}},\ }\href@noop {} {} (\bibinfo {year} {2022}),\ \Eprint
  {http://arxiv.org/abs/2212.00053} {arXiv:2212.00053 [cond-mat.str-el]}
  \BibitemShut {NoStop}%
\bibitem [{\citenamefont {Fu}\ \emph {et~al.}(2018)\citenamefont {Fu},
  \citenamefont {Knolle},\ and\ \citenamefont {Perkins}}]{fu_majorana_2018}%
  \BibitemOpen
  \bibfield  {author} {\bibinfo {author} {\bibfnamefont {J.}~\bibnamefont
  {Fu}}, \bibinfo {author} {\bibfnamefont {J.}~\bibnamefont {Knolle}}, \ and\
  \bibinfo {author} {\bibfnamefont {N.~B.}\ \bibnamefont {Perkins}},\ }\href
  {\doibase 10.1103/PhysRevB.97.115142} {\bibfield  {journal} {\bibinfo
  {journal} {Phys. Rev. B}\ }\textbf {\bibinfo {volume} {97}},\ \bibinfo
  {pages} {115142} (\bibinfo {year} {2018})}\BibitemShut {NoStop}%
\bibitem [{\citenamefont {Burnell}\ and\ \citenamefont
  {Nayak}(2011)}]{burnell11}%
  \BibitemOpen
  \bibfield  {author} {\bibinfo {author} {\bibfnamefont {F.~J.}\ \bibnamefont
  {Burnell}}\ and\ \bibinfo {author} {\bibfnamefont {C.}~\bibnamefont
  {Nayak}},\ }\href {\doibase 10.1103/PhysRevB.84.125125} {\bibfield  {journal}
  {\bibinfo  {journal} {Phys. Rev. B}\ }\textbf {\bibinfo {volume} {84}},\
  \bibinfo {pages} {125125} (\bibinfo {year} {2011})}\BibitemShut {NoStop}%
\bibitem [{\citenamefont {Kiselev}(2006)}]{doi:10.1142/S0217979206033310}%
  \BibitemOpen
  \bibfield  {author} {\bibinfo {author} {\bibfnamefont {M.~N.}\ \bibnamefont
  {Kiselev}},\ }\href {\doibase 10.1142/S0217979206033310} {\bibfield
  {journal} {\bibinfo  {journal} {International Journal of Modern Physics B}\
  }\textbf {\bibinfo {volume} {20}},\ \bibinfo {pages} {381} (\bibinfo {year}
  {2006})}\BibitemShut {NoStop}%
\end{thebibliography}
\end{document}